%Paper: hep-th/9505093
%From: BRUNELLI@urhep.pas.rochester.edu
%Date: Tue, 16 May 1995 10:18:30 -0500 (EST)

\magnification=1200
\baselineskip=13pt
\overfullrule=0pt
\tolerance=100000
\font\pq=cmr10 at 8truept

{\hfill \hbox{\vbox{\settabs 1\columns
\+ UR-1422 \cr
\+ ER-40685-871\cr
\+ hep-th/9505093\cr
}}}

\bigskip
\bigskip
\baselineskip=18pt

\centerline{\bf Supersymmetric Two Boson Equation, Its Reductions}
\centerline{\bf and the Nonstandard Supersymmetric KP Hierarchy}
\vfill

\centerline{J. C. Brunelli}
\medskip
\centerline{and}
\medskip
\centerline{Ashok Das}
\medskip
\medskip
\centerline{Department of Physics and Astronomy}
\centerline{University of Rochester}
\centerline{Rochester, NY 14627, USA}
\vfill

\centerline{\bf {Abstract}}

\medskip
\medskip

In this paper, we review various properties of the supersymmetric Two Boson
(sTB) system. We discuss the equation and its nonstandard Lax representation.
We construct the local conserved charges as well as the Hamiltoniam structures
of the system. We show how this system leads to various other known
supersymmetric integrable models under appropriate field redefinition. We
discuss the sTB and the supersymmetric nonlinear Schr\"odinger (sNLS) equations
as constrained, nonstandard supersymmetric Kadomtsev-Petviashvili (sKP)
systems and point out that the nonstandard sKP systems naturally unify all the
KP and mKP flows while leading to a new integrable supersymmetrization of the
KP equation. We construct the nonlocal conserved charges associated with the
sTB system and show that the algebra of charges corresponds to a graded, cubic
algebra. We also point out that the sTB system has a hidden supersymmetry
making it an $N=2$ extended supersymmetric system.
\vfill
\eject
\bigskip
\noindent {\bf 1. {Introduction}}
\medskip
Integrable models have been studied in the past from many different points
of view [1-3]. These models have a very rich structure and as
such deserve to be
studied in their own right. Recently, however, these models have become
quite relevant in the study of strings through the matrix models [4]. This has
generated additional interest, particularly, in the supersymmetric
integrable models since they are the ones which are likely to be
relevant in the study of superstrings [5,6].

While the bosonic integrable models have been studied quite extensively,
not much is known, in general, about the supersymmetric integrable models.
The most widely studied supersymmetric integrable system is
the supersymmetric KdV
(sKdV) system [7,8] and has many interesting properties. Another super
integrable model which has a very rich structure and which, in fact, leads
to the sKdV system upon reduction
is the supersymmetric KP equation of Manin-Radul [7]. While some of
its properties are known, a lot remains to be known about the properties
of this highly nontrivial system.

There is another supersymmetric integrable system that also has an equally rich
structure and is the subject of discussion of this paper. It is the
supersymmetric generalization of what is known as the Two Boson system [9].
This system, known as the supersymmetric Two Boson (sTB)
system [10] is rich in the
sense that it gives rise to many other supersymmetric integrable systems under
appropriate field redefinitions or reductions -- much like the sKP system.
Yet, its structure is simple enough to determine many of its properties
explicitly. In this paper, we review all the known properties of this system.

Our paper is organized as follows. In section 2, we discuss the Two Boson
system [9] and discuss briefly various known properties of this system. In
section 3, we construct the supersymmetric generalization of this system
(sTB) [10] and discuss its Lax representation as well as the local conserved
charges associated with it. In section 4, we generalize the three
Hamiltonian structures of the bosonic system to the superspace and
construct the recursion operator which relates different Hamiltonians of
the system. We point out that this is truly a bi-Hamiltonian super
integrable system (in the sense that there are two local, compatible
Hamiltonian
structures) -- in fact, the only one that we are aware of. In section
5, we discuss how the supersymmetric nonlinear Schr\"odinger equation
(sNLS), the sKdV equation and the smKdV equations can be obtained from
this system. The Hamiltonian structures of these systems can also be
obtained from those of the sTB system and we indicate how this is carried
out [11]. The Lax operator
for the sTB system gives rise to a scalar Lax operator
for the sNLS equation which shows that the sNLS system can be thought of
as a constrained, nonstandard supersymmetric KP system (sKP)
and this is discussed in section 6. This constrained,
nonstandard  sKP system unifies all the KP and mKP flows into a
single equation and leads to a new supersymmetric form of the KP equation
that is integrable [12]. In section 7, we construct the conserved nonlocal
charges associated with the sTB system and show that their algebra is an
interesting, cubic graded algebra. This also shows that the sTB equation
has a second supersymmetry which is not manifest making it an $N=2$
extended supersymmetric system [13]. We present our conclusions in section 8.
In particular, we clarify the relation between some recently proposed
equations with that of ours [14-17]. We
compile some of the necessary technical
formulae in the Appendix.

\bigskip
\noindent {\bf 2. {Two Boson Hierarchy}}
\medskip

The  dispersive generalization of the long water wave
equation [9,18,19] in a narrow channel has the form
$$
\eqalign{
{\partial u \over \partial  t} &= (2h + u^2 - \alpha
 u^\prime)^\prime\cr
\noalign{\vskip 4pt}%
{\partial  h \over \partial  t} &= (2uh + \alpha h^\prime)^\prime\cr
}
\eqno(2.1)
$$
where $u(x,t)$ and $h(x,t)$ can be thought of as the horizontal velocity
and the height (vertical displacement from the mean depth), respectively,
of the free surface ($\alpha$ is  an arbitrary parameter)
and a prime denotes a derivative with respect to $x$. This system of
equations is integrable [19] and has a tri-Hamiltonian structure [9].
It has
a nonstandard Lax representation and reduces to various known integrable
systems with appropriate identification [9]. For example, with the
identification
$$
\eqalign{
\alpha &= 1\cr
u &= - q^\prime / q\cr
h &= q \overline  q\cr
}\eqno(2.2)
$$
we get the nonlinear Schr\"odinger equation [20-23], with $\alpha =-1$,
$h=0$, (2.1) gives the Burgers' equation while for $\alpha=0$ we obtain
Benney's equation which is the standard long wave equation. The KdV  and the
mKdV equations can also be obtained from the hierarchy associated with this
integrable equation. Thus, (2.1) in different limits,
describes the long wave equation (Benney's equation, KdV equation) as well as
the short wave equation (the nonlinear Schr\"{o}dinger equation) [24]. This
equation, therefore, has a very rich structure much like the KP equation in two
dimensions. In fact, as we will see, it has a lot in common with the KP
equation which has been studied extensively in the literature.

{}From now on we will choose $\alpha=1$ and we
will make the identification
$u=J_0$ and $h=J_1$, which is used in the Two Boson formulation of this system
[20,25]. The equations in (2.1) then take the form
$$
\eqalign{
{\partial  J_0 \over \partial  t} &= (2 J_1 + J_0^2 -
J^\prime_0 )^\prime\cr
\noalign{\vskip 4pt}%
{\partial  J_1 \over \partial  t} &= (2 J_0 J_1 +
J^\prime_1 )^\prime\cr
}\eqno(2.3)
$$
In this form, the equations are commonly referred to as the Two Boson equation
(TB) and this is the form of the equations that we will use in our discussions.
It is easy to see that the variables of the system can be assigned the
following canonical dimensions.
$$
[x] = -1 \qquad [t] =-2 \qquad [J_0] = 1 \qquad [J_1 ] = 2 \eqno(2.4)
$$

The system of equations (2.3) can be written in the form of a Lax
equation [9,25]
$$
{\partial  L \over \partial  t} = \left[ L, \left( L^2 \right)_{\geq 1}
\right] \eqno(2.5)
$$
where the Lax operator has the form
$$
L =  \partial  - J_0 + \partial^{-1} J_1 \eqno(2.6)
$$
and $()_{\geq 1}$ refers to the differential part of a
pseudo-differential operator. This is conventionally called a nonstandard
representation of the Lax equation. We note that canonical dimension of the Lax
operator is given by
$$
[L] = 1\eqno(2.7)
$$
The conserved quantities of
the system (which are in involution) are given by
$$
H_n = \ {\rm Tr}\ L^n = \int dx\ {\rm Res}\ L^n \qquad \qquad
n = 1,2,3,\dots \eqno(2.8)
$$
where "Res" stands for the coefficient of the $\partial^{-1}$ term in the
pseudo-differential operator and the first few Hamiltonians have the explicit
forms
$$
H_1 = \int dx\ J_1 \qquad H_2 = \int dx\ J_0 J_1 \qquad
H_3= \int dx\ \left( J^2_1 - J_0^\prime J_1 + J_1 J^2_0 \right) \eqno(2.9)
$$

Given a Lax equation, one can obtain the Hamiltonian structures associated with
the dynamical equations through the Gelfand-Dikii brackets in a
straightforward manner at least for standard Lax representations [3,26].
However, the present
system corresponds to a nonstandard Lax representation [27]
and consequently, the
standard definitions of the Gelfand-Dikii brackets need to be generalized in
this case. There are two possible, but equivalent, generalizations which lead
to the correct Hamiltonian structures of the system [25]. First,
for the Lax operator
$$
L=\partial + {\overline J}_0 + \partial^{-1} {\overline J}_1
\qquad\qquad \big(\, {\overline J}_0 = - J_0 \quad {\overline J}_1 = J_1 \big)
\eqno(2.10)
$$
the standard definition of a dual
$$
Q = q_0 + q_1 \partial^{-1}\eqno(2.11)
$$
leads to a linear functional of
$\left(\,{\overline J}_0 ,{\overline J}_1 \right)$ of the form
$$
F_Q (L) = \ {\rm Tr}\ LQ = \sum^1_{i=0}
\int dx\ q_{1-i} (x)
{\overline J}_i (x)
\eqno(2.12)
$$
We can modify the definition of the dual as
$$
{\overline Q} = Q + q_{-1}\partial \eqno(2.13)
$$
with the constraint (The structure of the Lax equations leads to such
constraints [25].)
$$
q_{-1} J_1 = q_1\eqno(2.14)
$$
which would lead to the Hamiltonian structures from the standard definition of
Gelfand-Dikii brackets as
$$
\eqalign{
\big\{ F_{\overline Q} (L), F_{\overline V} (L) \big\}_1 =\ &
{\rm Tr}\ L [{\overline Q},{\overline V}]\cr
\noalign{\vskip 4pt}%
\big\{ F_{\overline Q} (L), F_{\overline V} (L)\big\}_2 = \ &{\rm Tr}\ ( (
L({\overline V}L)_+ - (L{\overline V})_+ L){\overline Q})\cr
}
\eqno(2.15)
$$
Equivalently, we can keep the definition of the dual in (2.11) and modify the
definition of the Gelfand-Dikii brackets as
$$
\eqalign{
\big\{ F_Q (L), F_V (L) \big\}_1 =\ &{\rm Tr}\ L [Q,V]\cr
\noalign{\vskip 4pt}%
\big\{ F_Q (L), F_V (L)\big\}_2 = \ &{\rm Tr}\ ( (
L(VL)_+ - (LV)_+ L)Q )\cr
&+ \int dx\ \bigg[ \bigg( \int^x \ {\rm Res}\ [Q,L] \bigg) \ {\rm Res}\
[V,L]\cr
&+\ {\rm Res}\ [Q,L] \ {\rm Res}\ \big( \partial^{-1} LV \big) -\ {\rm
 Res}\ [V,L] \ {\rm Res}\
\big( \partial^{-1} LQ \big) \bigg]\cr
}\eqno(2.16)
$$

Either way, the first two Hamiltonian structures of the system
can be derived from (2.15) or (2.16). The first structure turns out to be
simple
$$
\pmatrix{\big\{ J_0 (x) , J_0 (y) \big\}_1 &\big\{J_0 (x), J_1 (y) \big\}_1 \cr
\noalign{\vskip 12pt}%
\big\{ J_1 (x) , J_0 (y) \big\}_1&\big\{ J_1 (x) , J_1 (y) \big\}_1 }=
\pmatrix{0 &\partial\cr
\noalign{\vskip 5pt}%
\partial&0}\delta(x-y)={\cal D}_1\delta(x-y)
\eqno(2.17)
$$
where $\partial\equiv{\partial\over\partial x}$. Similarly, the second
structure is easily obtained to be
$$
\pmatrix{\big\{ J_0 (x) , J_0 (y) \big\}_2 &\big\{J_0 (x), J_1 (y) \big\}_2 \cr
\noalign{\vskip 12pt}%
\big\{ J_1 (x) , J_0 (y) \big\}_2&\big\{ J_1 (x) , J_1 (y) \big\}_2 }=
\pmatrix{2\partial &\partial J_0 - \partial^2\cr
\noalign{\vskip 12pt}%
J_0\partial+\partial^2&J_1\partial+\partial J_1}\delta(x-y)=
{\cal D}_2\delta(x-y)
\eqno(2.18)
$$
Note that if we introduce the recursion operator
$$
R={\cal D}_2{\cal D}_1^{-1}\eqno(2.19)
$$
we can obtain a third Hamiltonian structure for the system as

$$
{\cal D}_3=R\,{\cal D}_2\eqno(2.20)
$$
whose explicit form is complicated and not very interesting, except for the
fact that it is local. But we note that
eqs. (2.17), (2.18) and (2.20) define the three Hamiltonian structures of the
Two Boson equation (hierarchy) as
$$
\partial_t\pmatrix{J_0\cr
\noalign{\vskip 10pt}%
J_1}={\cal D}_1
\pmatrix{{\delta H_{3}\over\delta J_0}\cr
\noalign{\vskip 10pt}%
{\delta H_{3}\over\delta J_1}}=
{\cal D}_2
\pmatrix{{\delta H_{2}\over\delta J_0}\cr
\noalign{\vskip 10pt}%
{\delta H_{2}\over\delta J_1}}=
{\cal D}_3
\pmatrix{{\delta H_{1}\over\delta J_0}\cr
\noalign{\vskip 10pt}%
{\delta H_{1}\over\delta J_1}}
\eqno(2.21)
$$
making it a tri-Hamiltonian system [9]. It is worth emphasizing here that
integrable systems are in general multi-Hamiltonian [1-3],
but what is special about the Two Boson equation is that the
three distinct Hamiltonian structures we have just derived are local.
It is also easy to check that
the conserved charges of the system are recursively related by the recursion
operator as
$$
\pmatrix{{\delta H_{n+1}\over\delta J_0}\cr
\noalign{\vskip 10pt}%
{\delta H_{n+1}\over\delta J_1}}=R^\dagger
\pmatrix{{\delta H_{n}\over\delta J_0}\cr
\noalign{\vskip 10pt}%
{\delta H_{n}\over\delta J_1}}\eqno(2.22)
$$

We also note here for completeness that, if we define
$$
\eqalign{J(x) &= J_0 (x)\cr
T(x) &= J_1 (x) - {1 \over 2}\ J^\prime_0 (x)\cr}
\eqno(2.23)
$$
then, the second Hamiltonian structure in (2.18), in these variables,
has the form [25]
$$
\eqalign{\{ J(x) , J(y) \}_2 &= 2 \partial_x \delta (x-y)\cr
\{ T(x) , J(y) \}_2 &= J(x) \partial_x \delta (x-y)\cr
\{ T(x) , T(y)\}_2 &= (T (x) + T(y)) \partial_x \delta (x-y) +
{1 \over 2}\ \partial^3_x \delta (x-y)\cr}
\eqno(2.24)
$$
which is a Virasoro-Kac-Moody algebra for a $U(1)$ current [28,29]
(also known as an affine algebra [30]).
The second Hamiltonian structure, in (2.18) or (2.24), as we will see in
section 8, corresponds to the bosonic limit of the $N=2$ twisted superconformal
algebra [31].

As we have already noted earlier, with the field identifications [20-23]
$$
\eqalign{
J_0=&-{q'\over q}=-(\ln q )'\cr
J_1=&{\bar q}q
}\eqno(2.25)
$$
we obtain from  (2.3), the nonlinear Schr\"odinger equation
$$
\eqalign{
{\partial q\over\partial t}=&-(q''+2({\bar q}q)q)\cr
\noalign{\vskip 4pt}%
{\partial {\bar q}\over\partial t}=&\,\,{\bar q}''+2({\bar q}q){\bar q}\cr
}\eqno(2.26)
$$
The Lax operator in (2.6), with the identification in (2.25),
factorizes and has the form
$$
\eqalign{
L=&\partial+{q'\over q}+\partial^{-1}{\bar q}q \cr
 =&G{\widetilde L}G^{-1}\cr
}\eqno(2.27)
$$
where
$$
\eqalign{
G=&q^{-1}\cr
{\widetilde L}=&\partial+q\partial^{-1}{\bar q}\cr
}\eqno(2.28)
$$
We say that the two Lax operators, $L$ and ${\widetilde L}$, are related
through
a gauge transformation [20,23,32].  The new Lax
operator, ${\widetilde L}$, allows us
to write the nonlinear Schr\"odinger equation in the standard Lax
representation
$$
{\partial {\widetilde L}\over\partial t}=\left[{\widetilde L},
({\widetilde L}^2)_{+}\right]\eqno(2.29)
$$
providing a scalar Lax operator for the nonlinear Schr\"odinger equation
which puts it in the same footing as the other integrable equations such as
the KdV equation. The Lax operator, $\widetilde L$, in (2.28) can also be
written as
$$
\eqalign{
{\widetilde L}=&\partial +q{\bar q}\,\partial^{-1}-q{\bar q}'\partial^{-2}+
q{\bar q}''\partial^{-3}+\cdots\cr
   =&\partial +\sum_{n=0}^\infty u_n\partial^{-n-1}\cr
}\eqno(2.30)
$$
with
$$
u_n=(-1)^nq{\bar q}^{(n)}\eqno(2.31)
$$
where $f^{(n)}$ is the $n$th derivative of $f$ with respect to $x$. The form of
$\widetilde L$ in (2.30) is the same as that of a KP system with coefficient
functions constrained by (2.31). This allows us to think of the nonlinear
Schr\"odinger equation as a constrained  KP system [20,22,27].

Some observations are in order here for the discussion of the supersymmetric
generalization of this system which we are going to take up in the coming
sections. Let us note that given $\widetilde L$, we can define
its  formal adjoint [33]
$$
{\cal L}={\widetilde L}^*=-(\partial+{\bar q}\partial^{-1}q)\eqno(2.32)
$$
and the standard Lax equation
$$
{\partial{\cal L}\over\partial t}=\left[\left({\cal L}^2\right)_
+,{\cal L}\right]
\eqno(2.33)
$$
also gives the nonlinear Schr\"odinger equation. Furthermore, with the
identification
$$
{\bar q}=q\eqno(2.34)
$$
the standard Lax equation
$$
{\partial {\widetilde L}\over\partial t}=\left[{\widetilde L},
({\widetilde L}^3)_{+}\right]\eqno(2.35)
$$
leads to the mKdV equation,
$$
{\partial q\over\partial t}=-(q'''+6q^2q')\eqno(2.36)
$$
The Lax operator, $\cal L$, in (2.32) with the same identification
as in (2.34), becomes
$$
{\cal L}=-{\widetilde L}\eqno(2.37)
$$
and also gives the mKdV equation from
$$
{\partial{\cal L}\over\partial t}=\left[\left({\cal L}^3\right)_+
,{\cal L}\right]
\eqno(2.38)
$$
This shows how the mKdV equation can be embedded into the nonlinear
Schr\"odinger equation and from our discussion above, it is clear that both
the Lax operator and its formal adjoint yield equivalent results. We will see
later that this equivalence no longer holds in the
supersymmetric generalization of this system.
\bigskip
\noindent {\bf 3. {Supersymmetric Two Boson Hierarchy}}
\medskip

The easiest way to derive the supersymmetric generalization of the Two Boson
equation is to go to the superspace [10]. Let $z = (x,\theta)$ define the
coordinates of the superspace with $\theta$ representing the Grassmann
coordinate and
$$
D = {\partial  \over \partial  \theta} + \theta {\partial
 \over \partial  x}
\eqno(3.1)
$$
representing the supercovariant derivative. We see from  (3.1) that
$D^2 = \partial$,  and this leads to
$$
[\theta] = - {1 \over 2}
\eqno(3.2)
$$
Let us now introduce two fermionic superfields
$$
\eqalign{\Phi_0 &= \psi_0 + \theta J_0\cr
\Phi_1 &= \psi_1 + \theta J_1 \cr
}
\eqno(3.3)
$$
which have the following canonical dimensions
$$
\eqalign{
[\Phi_0] &= [ \psi_0] = {1 \over 2}\cr
\noalign{\vskip 4pt}%
[\Phi_1] &= [ \psi_1] = {3 \over 2}\cr
}\eqno(3.4)
$$
This would correspond to a simple extension of the variables of our bosonic
system to an $N=1$ supersymmetric case.

With the superfields in (3.3) as our basic variables, we can now write the most
general, local dynamical equations in superspace which are
consistent with all the canonical dimensions and which reduce to (2.3)
in the bosonic limit. These are easily checked to be
$$
\eqalign{
{\partial  \Phi_0 \over \partial  t} = &-(D^4 \Phi_0) + 2(D \Phi_0)
(D^2 \Phi_0) + 2(D^2 \Phi_1)\cr
&+ a_1 D(\Phi_0 (D^2 \Phi_0)) + a_2 D (\Phi_0 \Phi_1)\cr
\noalign{\vskip 4pt}%
{\partial  \Phi_1 \over \partial  t} = &(D^4 \Phi_1) + b_1 D( ( D^2
\Phi_1 ) \Phi_0) + 2(D^2 \Phi_1)(D \Phi_0 ) - b_2 D(\Phi_1
(D^2 \Phi_0))\cr
& +2(D \Phi_1 )(D^2 \Phi_0) + b_3 \Phi_1 \Phi_0(D^2 \Phi_0 ) + b_4 D(\Phi_1
\Phi_0) (D \Phi_0)\cr
&+ b_5 D(\Phi_0 (D^4 \Phi_0)) +b_6 D(\Phi_0 (D^2 \Phi_0)) (D \Phi_0)\cr
}
\eqno(3.5)
$$
where $a_1$, $a_2$, $b_1$, $b_2$, $b_3$, $b_4$, $b_5$ and $b_6$ are
arbitrary parameters. An integrable system is likely to have a Lax
representation and, therefore,  we look for a Lax
representation for the supersymmetric equations (3.5). It is easy to see that
a consistent Lax equation can be defined if we choose, as the Lax operator,
the pseudo super-differential operator
$$
L = D^2 + \alpha (D \Phi_0) + \beta D^{-1} \Phi_1
\eqno(3.6)
$$
where $\alpha$, $\beta$ are arbitrary parameters and $D^{-1}=\partial^{-1}D$ is
the inverse of the covariant derivative operator. (Note that for $\alpha=-1$
and $\beta = 1$ (3.6) reduces to (2.6) in the bosonic limit.)
The nonstandard Lax equation
$$
{\partial  L \over \partial  t} = \left[ L, (L^2)_{\geq 1}
\right]\eqno(3.7)
$$
in this case, gives
$$
\eqalign{{\partial  \Phi_0 \over \partial  t} &= -(D^4 \Phi_0) - 2 \alpha (D
\Phi_0)
(D^2 \Phi_0) - {2 \beta \over \alpha} \ (D^2 \Phi_1)\cr
\noalign{\vskip 4pt}%
{\partial  \Phi_1 \over \partial  t} &= (D^4 \Phi_1) - 2 \alpha D^2
((D \Phi_0 )\Phi_1)\cr}
\eqno(3.8)
$$
Comparing with (3.5), we conclude that both the equations are compatible only
if $\alpha = -1$ and $\beta = 1$ so that
$$
L = D^2 - (D \Phi_0) + D^{-1} \Phi_1
\eqno(3.9)
$$
and the most general supersymmetric extension of the dynamical equations in
(2.3) which is integrable is given by [10]
$$
\eqalign{{\partial  \Phi_0 \over \partial  t} &=
 - (D^4 \Phi_0) + (D(D\Phi_0)^2)+ 2(D^2 \Phi_1)\cr
\noalign{\vskip 4pt}%
{\partial  \Phi_1 \over \partial  t} &=
 (D^4 \Phi_1) + 2(D^2((D \Phi_0) \Phi_1))\cr
}
\eqno(3.10)
$$

In components, (3.10) gives rise to the following system of interacting
equations
$$
\eqalign{
{\partial  J_0 \over \partial  t} &= ( 2J_1 + J_0^2 - J_0^\prime)^\prime\cr
\noalign{\vskip 4pt}%
{\partial  \psi_0 \over \partial  t} &= 2 \psi_1^\prime + 2 \psi_0^\prime
J_0 - \psi_0^{\prime \prime}\cr
\noalign{\vskip 4pt}%
{\partial  J_1 \over \partial  t} &= ( 2J_0 J_1 + J_1^\prime
 + 2 \psi_0^\prime \psi_1)^\prime\cr
\noalign{\vskip 4pt}%
{\partial  \psi_1 \over \partial  t} &= ( 2 \psi_1
J_0 + \psi_1^\prime)^\prime\cr
}
\eqno(3.11)
$$
And it is straightforward to check that it is invariant under the
supersymmetry transformations
$$
\eqalign{
\delta  J_0 &= \epsilon \psi_0^\prime\cr
\delta  J_1 &= \epsilon \psi_1^\prime\cr
\delta \psi_0 &= \epsilon J_0\cr
\delta  \psi_1 &= \epsilon J_1\cr
}
\eqno(3.12)
$$
where $\epsilon$ is a constant Grassmann parameter. This, therefore, appears to
be the most general $N=1$ supersymmetric extension of the Two Boson equation
which is integrable. However, as we will see later, there are hidden symmetries
in this system [13]. In particular, among other things, we will see
that there is a second supersymmetry [13,15,16] associated with this system.

Equation (3.10) or (3.11) define an integrable system and therefore, there are
an infinite number of local conserved charges which can be
obtained in the standard manner to be
$$
Q_n=\hbox{sTr}\,L^n=\int dz\,\hbox{sRes}\,L^n\qquad n=1,2,\dots \eqno(3.13)
$$
where ``sRes'' stands for the super residue which is defined to be the
coefficient of the $D^{-1}$ term in the pseudo super-differential operator
with $D^{-1}$ at the right. The constancy of these charges under the flow of
(3.10) can, of course, be directly checked, but it also follows easily from the
structure of the Lax equation in (3.7). The first few charges have the explicit
form
$$
\eqalign{
Q_1=&-\int dz\, \Phi_1\cr
Q_2=&2\int dz\, (D\Phi_0)\Phi_1\cr
Q_3=&3\int dz\,\Bigl[(D^3\Phi_0)-(D\Phi_1)-(D\Phi_0)^2\Bigr]\Phi_1\cr
Q_4=&2\int
dz\,\Bigl[2(D^5\Phi_0)+2(D\Phi_0)^3+6(D\Phi_0)(D\Phi_1)-
3\left(D^2(D\Phi_0)^2\right)
\Bigr]\Phi_1\cr
}\eqno(3.14)
$$
We observe
that $[Q_n]=n$, they are bosonic and are invariant under the supersymmetry
transformations given by  (3.12). In fact,
since the supersymmetry transformations (3.12) can be thought of as
arising from a translation of the Grassmann
coordinate in the superspace, and since these charges are given as superspace
integrals of local functions, supersymmetry of these charges is manifest.

We also note here that associated with the supersymmetric Two Boson equation is
a hierarchy of supersymmetric integrable equations given by
$$
{\partial  L \over \partial  t_n} = \left[ L, (L^n)_{\geq 1}
\right]\eqno(3.15)
$$
with $L$ given in eq. (3.9). All the equations in (3.15) share the same
conserved charges and define the supersymmetric Two Boson (sTB) hierarchy. In
the next section, we will derive the Hamiltonian structures associated with
this hierarchy.
\bigskip
\noindent {\bf 4. {Hamiltonian Structures for the sTB}}
\medskip
As we have noted earlier, given a Lax equation, we can obtain the Hamiltonian
structures associated with a dynamical system from the
Gelfand-Dikii brackets [3,26].
The sTB hierarchy, as we have seen, has a nonstandard Lax representation. While
the generalization of the Gelfand-Dikii brackets for the bosonic nonstandard
Lax equations was obtained in ref. [25], the extension
of these brackets to  superspace is
technically more difficult and not yet understood. Therefore, we will construct
the brackets by supersymmetrizing the bosonic Hamiltonian structures
(2.17), (2.18) and (2.20) in a direct way [10,13].

Defining the Hamiltonians of the sTB system as
$$
H_n={(-1)^{n+1}\over n}Q_n\eqno(4.1)
$$
we note that the sTB equation (3.10) can be written as a
Hamiltonian system with three Hamiltonian structures of the form [10,13]
$$
\partial_t\pmatrix{\Phi_0\cr
\noalign{\vskip 10pt}%
\Phi_1}
={\cal D}_1
\pmatrix{{\delta H_{3}\over\delta\Phi_0}\cr
\noalign{\vskip 10pt}%
{\delta H_{3}\over\delta\Phi_1}}=
{\cal D}_2
\pmatrix{{\delta H_{2}\over\delta\Phi_0}\cr
\noalign{\vskip 10pt}%
{\delta H_{2}\over\delta\Phi_1}}=
{\cal D}_3
\pmatrix{{\delta H_{1}\over\delta\Phi_0}\cr
\noalign{\vskip 10pt}%
{\delta H_{1}\over\delta\Phi_1}}
\eqno(4.2)
$$
where  the first structure has the local form
$$
{\cal D}_1=\pmatrix{0 & -D\cr
\noalign{\vskip 5pt}%
-D & 0}\eqno(4.3)
$$
This can also be written as a matrix in the component space as
$$
{\cal D}_1=\pmatrix{0 & \partial & 0 & 0\cr
\noalign{\vskip 5pt}%
\partial & 0 & 0 & 0\cr
\noalign{\vskip 5pt}%
0 & 0 & 0 & -1\cr
\noalign{\vskip 5pt}%
0 & 0 & -1 & 0}\eqno(4.4)
$$
and yields the following nonvanishing Poisson brackets in components
$$
\eqalign{
\{\psi_0(x),\psi_1(y)\}_1=&-\delta(x-y)\cr
\noalign{\vskip 3pt}%
\{J_0(x),J_1(y)\}_1=&\delta'(x-y)\cr
}\eqno(4.5)
$$
The second Hamiltonian structure of the system is given by [13]
$$
{\cal D}_2=\pmatrix{-2D-2D^{-1}\Phi_1D^{-1}+D^{-1}(D^2\Phi_0)D^{-1}&
D^3-D(D\Phi_0)+D^{-1}\Phi_1D\cr
\noalign{\vskip 20pt}%
-D^3-(D\Phi_0)D-D\Phi_1D^{-1}&-\Phi_1D^2-D^2\Phi_1}\eqno(4.6)
$$
and can also be written as a matrix in the component space as
$$
{\cal D}_2 = \pmatrix{2\partial & \partial J_0-\partial^2 &
(\partial\psi_0)\partial^{-1}-2\psi_1\partial^{-1} & \psi_1\cr
\noalign{\vskip 12pt}%
J_0\partial+\partial^2 & \partial J_1+J_1\partial &
-(\partial\psi_0)-\partial\psi_1\partial^{-1} &
\partial\psi_1+\psi_1\partial\cr
\noalign{\vskip 12pt}%
\partial^{-1}(\partial\psi_0)-2\partial^{-1}\psi_1 &
(\partial\psi_0)+\partial^{-1}\psi_1\partial & \partial^{-1}((\partial
J_0)-2J_1)\partial^{-1}-2 & \partial^{-1} J_1-J_0+\partial\cr
\noalign{\vskip 12pt}%
-\psi_1 & \partial\psi_1+\psi_1\partial & -J_1\partial^{-1}-J_0-\partial & 0}
\eqno(4.7)
$$
which gives the nonvanishing  Poisson brackets in the components of the form
$$
\eqalign{
\{\psi_0(x),\psi_0(y)\}_2=&
\Bigl(\partial^{-1}J_0'\bigl(\partial^{-1}\delta(x-y)\bigr)\Bigr)
-2\Bigl(\partial^{-1}J_1\bigl(\partial^{-1}\delta(x-y)\bigr)\Bigr)
-2\delta(x-y)\cr
\noalign{\vskip 3pt}%
\{\psi_0(x),J_0(y)\}_2=
&\Bigl(\partial^{-1}\psi'_0\delta(x-y)\Bigr)
-2\Bigl(\partial^{-1}\psi_1\delta(x-y)\Bigr)\cr
\noalign{\vskip 3pt}%
\{J_0(x),J_0(y)\}_2=&2\delta'(x-y)\cr
\noalign{\vskip 3pt}%
\{\psi_0(x),\psi_1(y)\}_2=&
\Bigl(\partial^{-1}J_1\delta(x-y)\Bigr)-J_0\delta(x-y)+\delta'(x-y)\cr
\noalign{\vskip 3pt}%
\{\psi_0(x),J_1(y)\}_2=
&\psi'_0\delta(x-y)+\Bigl(\partial^{-1}\psi_1\delta'(x-y)\Bigr)\cr
\noalign{\vskip 3pt}%
\{J_0(x),\psi_1(y)\}_2=&\psi_1\delta(x-y)\cr
\noalign{\vskip 3pt}%
\{J_0(x),J_1(y)\}_2=&(J_0\delta(x-y))'-\delta''(x-y)\cr
\noalign{\vskip 3pt}%
\{\psi_1(x),J_1(y)\}_2=&2\psi_1\delta'(x-y)+\psi_1'\delta(x-y)\cr
\noalign{\vskip 3pt}%
\{J_1(x),J_1(y)\}_2=&J_1'\delta(x-y)+2J_1\delta'(x-y)\cr
\noalign{\vskip 3pt}%
}\eqno(4.8)
$$
Introducing the recursion operator
$$
R={\cal D}_2{\cal D}_1^{-1}\eqno(4.9)
$$
the third Hamiltonian structure can be written as
$$
{\cal D}_3=R\,{\cal D}_2\eqno(4.10)
$$
whose explicit form is complicated and uninteresting.

While the skew-symmetry character of the Hamiltonian structures in (4.3), (4.6)
and (4.10) are easy to see, proof of the Jacobi identity for these structures,
at least, for  ${\cal D}_2$ and ${\cal D}_3$ is not at all obvious. Neither is
the bi-Hamiltonian character of this system (which follows from the
compatibility of the Hamiltonian structures or the fact that a linear
superposition of the structures is also a Hamiltonian structure). We will now
sketch how these can be checked for the Hamiltonian structures of our system
using prolongation methods described in [34] and generalized to
the supersymmetric systems in [35]. Let us define a two component
column matrix of bosonic superfields as
$$
{\vec\Omega}=\pmatrix{
\Omega_0\cr
\noalign{\vskip 10pt}%
\Omega_1}\eqno(4.11)
$$
and construct the bivector $\Theta_{\cal D}$ associated with any Hamiltonian
structure $\cal D$ as
$$
\Theta_{\cal D}={1\over 2}\sum_{\alpha,\beta}\int dz\,
\left(({\cal D})_{\alpha\beta}\Omega_\beta\right)
\wedge\Omega_\alpha\qquad
\alpha,\beta=0,1\eqno(4.12)
$$
Then, a necessary and sufficient condition for ${\cal D}$ to define a
Hamiltonian structure is that the prolongation of this bivector must vanish (We
refer interested readers to refs. [34] and [35] for details.)
$$
\hbox{\bf pr}\,{\vec v}_{{\cal D}{\vec\Omega}}
(\Theta_{{\cal D}}) = 0\eqno(4.13)
$$

For ${\cal D}_2$, the second Hamiltonian structure of sTB, it is
straightforward
to show that
$$
\eqalignno{
\hbox{\bf pr}\,{\vec v}_{{\cal D}_2{\vec\Omega}}(\Phi_0)=&
-2(D\Omega_0)-2\left(D^{-1}(\Phi_1(D^{-1}\Omega_0))\right)+
\left(D^{-1}(D^2\Omega_0)(D^{-1}\Omega_0)\right)&\cr
&+(D^3\Omega_1)-(D^2\Phi_0)\Omega_1-(D\Phi_0)(D\Omega_1)+
\left(D^{-1}(\Phi_1(D\Omega_1))\right)&(4.14a)\cr
\noalign{\vskip 10pt}%
\hbox{\bf pr}\,{\vec v}_{{\cal D}_2{\vec\Omega}}(\Phi_1)=&
-(D^3\Omega_0)-(D\Phi_0)(D\Omega_0)-(D\Phi_1)(D^{-1}\Omega_0)+\Phi_1\Omega_0\cr
&-2\Phi_1(D^2\Omega_1)-(D^2\Phi_1)\Omega_1&(4.14b)\cr
}
$$
Using this, the prolongation of the bivector (4.12) for the second structure
is easily seen to vanish
$$
\hbox{\bf pr}\,{\vec v}_{{\cal D}_2{\vec\Omega}}
(\Theta_{{\cal D}_2}) = 0\eqno(4.15)
$$
implying that ${\cal D}_2$ satisfies the Jacobi identity. Also, since
${\cal D}_1$ is field independent, it follows directly that
$$
\hbox{\bf pr}\,{\vec v}_{{\cal D}_1{\vec\Omega}}
(\Theta_{{\cal D}_1}) = 0\eqno(4.16)
$$
It can be shown similarly that ${\cal D}_1$ and ${\cal D}_2$ are compatible,
in that if we define
$$
{\cal D}={\cal D}_2+\alpha{\cal D}_1\eqno(4.17)
$$
where $\alpha$ is an arbitrary constant, then
$$
\hbox{\bf pr}\,{\vec v}_{{\cal D}{\vec\Omega}}(\Theta_{{\cal D}})=
\hbox{\bf pr}\,{\vec v}_{{\cal D}{\vec\Omega}}(\Theta_{{\cal D}_2})+
\alpha\hbox{\bf pr}\,{\vec v}_{{\cal D}{\vec\Omega}}
(\Theta_{{\cal D}_1})=0\eqno(4.18)
$$
showing that any arbitrary linear superposition of the first and the second
Hamiltonian structures is also a Hamiltonian structure. In a similar
manner, it can be shown that the third structure, ${\cal D}_3$, is also
Hamiltonian and that the supersymmetric Two Boson equation is a
tri-Hamiltonian system much like its bosonic counterpart.

Several comments are in order here. First, the two higher structures,
${\cal D}_2$ and ${\cal D}_3$, are not local unlike the bosonic equation.
As we will show in sec. 8, the second Hamiltonian structure really becomes
local and coincides with the $N=2$ twisted superconformal algebra [31] in a
different basis (in variables which are linearly related). This,
therefore, represents the only supersymmetric system that we
know of which has two local Hamiltonian structures and in this sense
represents a truly bi-Hamiltonian system. We would also like to emphasize
here that even though the
Hamiltonian structures obtained in ref. [10] (through a
naive supersymmetrization of the bosonic Hamiltonian structures)
give the correct dynamical
equations as well as the correct recursion relations between the first few
conserved charges, they fail to satisfy Jacobi identity. The structures in
(4.3), (4.6) and (4.10) represent the true Hamiltonian structures of
the system.

Finally, we note here that the recursion operator defined in (4.9) from
the Hamiltonian structures, relates the conserved Hamiltonians of the
system recursively as
$$
\pmatrix{{\delta H_{n+1}\over\delta\Phi_0}\cr
\noalign{\vskip 10pt}%
{\delta H_{n+1}\over\delta\Phi_1}}=R^\dagger
\pmatrix{{\delta H_{n}\over\delta\Phi_0}\cr
\noalign{\vskip 10pt}%
{\delta H_{n}\over\delta\Phi_1}}\eqno(4.19)
$$
This is sufficient to show that the conserved charges are in involution
(This can also be seen directly from the Lax equation.) leading to the
integrability of the system. We note here, for later use, the explicit
form of this operator
$$
R^\dagger=\pmatrix{D^2-D^{-1}(D^2\Phi_0)+(D\Phi_0)+\Phi_1D^{-1}&
2(D\Phi_1)-2\Phi_1D-D^{-1}(D^2\Phi_1)\cr
\noalign{\vskip 20pt}%
2+D^{-2}\Phi_1D^{-1}-D^{-2}(D^2\Phi_0)D^{-1}&
-D^2-D^{-2}(D\Phi_1)+(D\Phi_0)+D^{-1}\Phi_1
}\eqno(4.20)
$$
In the next section, we will show how this supersymmetric system reduces to
various other known supersymmetric integrable systems, much like its
bosonic counterpart.

\bigskip
\noindent {\bf 5. {Reductions of sTB}}
\medskip
\medskip
\noindent {\bf 5.1 {Reduction to Supersymmetric KdV}}
\medskip

We have pointed out in sec. 2 that the  Two Boson system reduces to
various known integrable systems. For instance, the KdV equation can be
embedded into the Two Boson system in a nonstandard Lax representation [9].
Let us first show that the sKdV [7,8] can,
similarly, be embedded into the sTB [11].

Let us consider the Lax operator for the sTB system, $L$,  given in (3.9).
It is straightforward to show from its structure that
$$
(L^3)_{\geq 1} = D^6 + 3 D \Phi_1 D^2 -3D^2(D\Phi_0)D^2+3(D\Phi_0)^2D^2
+6\Phi_1(D\Phi_0)D
\eqno(5.1)
$$
It can now be shown in a simple manner that the nonstandard Lax equation,
$$
{\partial   L \over \partial  t} = [ L, (L^3)_{\geq 1}]
\eqno(5.2)
$$
leads to the  dynamical equations
$$
\eqalignno{
{\partial\Phi_1\over\partial  t} =&
-(D^6\Phi_1 ) - 3 D^2 \biggl( \Phi_1 (D \Phi_0)^2+(D^2\Phi_1)(D\Phi_0)+
\Phi_1(D\Phi_1)\biggr)&(5.3a)\cr
{\partial\Phi_0\over\partial  t} =&
\!-\!(D^6\Phi_0)\!+\! 3 D\biggl(\! \Phi_1 (D^2
\Phi_0)-2(D\Phi_1)(D\Phi_0)-\!{1\over3}(D\Phi_0)^3\!+\!
(D\Phi_0)(D^3\Phi_0)\!\biggr)&(5.3b)\cr
}
$$
Given this, it is immediately clear that the identification
$$
\eqalign{\Phi_0 & =  0\cr
\Phi_1 & = \Phi} \eqno(5.4)
$$
gives rise to the sKdV equation
$$
{\partial\Phi\over\partial  t} =
-(D^6\Phi) - 3 D^2\left(\Phi(D\Phi)\right)\eqno(5.5)
$$
and that the sKdV equation is embedded in the sTB system as a nonstandard
Lax equation (5.2)  with the choice of the dynamical variables given in
(5.4). We also see from (4.1) that with the condition (5.4), the even
Hamiltonians vanish whereas the odd ones are the same as those for the sKdV
system.

The reduction in (5.4) imposes a constraint on the sTB system and the
Hamiltonian structures of the sKdV system can be obtained from those of the sTB
system through a Dirac procedure [36]. If ${\cal D}$ represents any one of the
Hamiltonian structures of the sTB system and $H$ the appropriate
odd Hamiltonian of the system, we have
$$
\partial_t\pmatrix{\Phi_0\cr
\noalign{\vskip 10pt}%
\Phi_1}
={\cal D}
\pmatrix{{\delta H\over\delta\Phi_0}\cr
\noalign{\vskip 10pt}%
{\delta H\over\delta\Phi_1}}
\eqno(5.6)
$$
Imposing the constraint (5.4) in (5.6), we obtain
$$
\partial_t\pmatrix{0\cr
\noalign{\vskip 10pt}%
\Phi}
=
\pmatrix{{\overline{\cal D}}_{11} & {\overline{\cal D}}_{12}\cr
\noalign{\vskip 5pt}%
{\overline{\cal D}}_{21}& {\overline{\cal D}}_{22}}
\pmatrix{{\overline{\delta H\over\delta\Phi_0}}=v_0\cr
\noalign{\vskip 10pt}%
{\overline{\delta H\over\delta\Phi_1}}=v_1}
\eqno(5.7)
$$
where $\overline{{\cal O}}$ denotes the quantities ${\cal O}$ calculated
with the constraint (5.4). From (5.7), we note that consistency requires
$$
v_0=-{\overline{\cal D}}_{11}^{-1}{\overline{\cal D}}_{12}v_1\eqno(5.8)
$$
and that we can write
$$
\partial_t\Phi ={\overline{\cal D}}_{21}v_0 + {\overline{\cal D}}_{22}v_1
={\cal D}^{\,\hbox{\pq sKdV}}v_1 \eqno(5.9)
$$
with ($v_1$ for odd Hamiltonians is the same as ${{\delta
H}\over{\delta\Phi}}$ for the sKdV system.)
$$
{\cal D}^{\,\hbox{\pq sKdV}}={\overline{\cal D}}_{22}-
{\overline{\cal D}}_{21}{\overline{\cal D}}_{11}^{-1}{\overline{\cal D}}_{12}
\eqno(5.10)
$$
We note now that if we use the second Hamiltonian structure (4.6),
we obtain
$$
({\overline{\cal D}}_2)^{-1}_{11}=-{1\over2}D(D^3+\Phi)^{-1}D\eqno(5.11)
$$
and (5.10) leads to the standard second Hamiltonian structure [8] of sKdV
$$
{\cal D}_2^{\,\hbox{\pq sKdV}}=-{1\over2}
(D^5+3\Phi D^2+(D\Phi)D+2(D^2\Phi))\eqno(5.12)
$$

To obtain the first Hamiltonian structure we should remember that
when $\Phi_0=0$, all the even charges in (4.1) vanish. Consequently,
the first Hamiltonian structure of the sKdV is
obtained through Dirac reduction from ${\cal D}_0$ and not from ${\cal D}_1$,
as would be naively expected.  From the recursion relations, we know that
$$
{\cal D}_0=R^{-1}{\cal D}_1\eqno(5.13)
$$
where $R$ is the recursion operator of the sTB  given in (4.9). It is easy to
show that
$$
{\overline R}^{-1}=\pmatrix{4D\left({\cal D}_2^{\,\hbox{\pq
sKdV}}\right)^{-1}D&
-2D\left({\cal D}_2^{\,\hbox{\pq sKdV}}\right)^{-1}D^3\cr
\noalign{\vskip 20pt}%
2D^3\left({\cal D}_2^{\,\hbox{\pq sKdV}}\right)^{-1}D&
\qquad2D^2(D^3+\Phi)^{-1}D^{-1}(\Phi D^2+D^2\Phi)
\left({\cal D}_2^{\,\hbox{\pq sKdV}}\right)^{-1}D^3
}\eqno(5.14)
$$
Using the Dirac reduction relation (5.10) for the Hamiltonian structure
(5.13), we obtain
$$
{\cal D}_1^{\,\hbox{\pq sKdV}}=({\overline{\cal D}}_0)_{22}-
({\overline{\cal D}}_0)_{21}({\overline{\cal D}}_0)^{-1}_{11}
({\overline{\cal D}}_0)_{12}\eqno(5.15)
$$
which takes the form
$$
{\cal D}_1^{\,\hbox{\pq sKdV}}=-2D^2(D^3+\Phi)^{-1}D^2 \eqno(5.16)
$$
This is the nonlocal structure for the sKdV obtained in refs.
[37-40], but
our derivation [11] also shows that this structure satisfies Jacobi identity
since the Hamiltonian structures of the sTB system do. (The proof of
Jacobi identity for the structure in (5.16) is nontrivial and, to the best
of our knowledge, has not been demonstrated directly.)

\bigskip
\noindent {\bf 5.2 {Reduction to Supersymmetric NLS}}
\medskip

As we have pointed out, the field redefinition (2.25) takes the two
boson equation to the nonlinear Schr\"odinger equation. Therefore, it
would seen natural to find a field redefinition that will take the sTB
equation to the sNLS equation. In fact with the field
redefinitions [10] (the only consistent ones and which reduces
to (2.25) in the bosonic limit)
$$
\eqalign{
\Phi_0 &= - \left(D \ln (DQ)\right) +
\left(D^{-1} (\overline Q Q)\right)\cr
%\noalign{\vskip 4pt}%
\Phi_1 &= - \overline Q ( DQ)\cr}
\eqno(5.17)
$$
in terms of fermionic superfields
$$
\eqalign{Q &= \psi + \theta q\cr
{\overline Q} &= {\overline\psi} + \theta {\overline q} \cr
}
\eqno(5.18)
$$
which are complex conjugate of each other, equations (3.10), after a
slightly involved derivation, reduce to
$$
\eqalign{{
\partial  Q \over \partial  t} &=
 -(D^4 Q) + 2\left(D((DQ){\overline Q})\right)Q\cr
\noalign{\vskip 4pt}%
{\partial  \overline Q \over \partial  t} &=
 (D^4 \overline Q ) - 2\left(D((D{\overline Q})Q)\right){\overline Q}\cr
}\eqno(5.19)
$$
This is the supersymmetric nonlinear Schr\"odinger equation that is integrable.
In fact, the supersymmetrization of the NLS equation [41,42] leads to the
following equations
$$
\eqalign{{
\partial  Q \over \partial  t} &=-(D^4 Q) + 2\alpha(D{\overline Q})(DQ)Q -
2\gamma{\overline Q}Q(D^2Q)+2(1-\alpha){\overline Q}(DQ)^2\cr
\noalign{\vskip 4pt}%
{\partial  \overline Q \over \partial  t} &= (D^4 \overline Q )-
2\alpha(DQ)(D{\overline Q}){\overline Q} +
2\gamma Q{\overline Q}(D^2{\overline Q})-2(1-\alpha)Q(D{\overline Q})^2\cr
}\eqno(5.20)
$$
However, it has been shown in [42] that various tests of integrability
hold for the system of equations (5.20) only for
$$
\alpha=-\gamma=1\eqno(5.21)
$$
suggesting that the system is integrable only for these values of
the parameters. Equation (5.19) is indeed the equation (5.20)
for the values of the parameter in
(5.21) and we have obtained it from the integrable sTB equation through a
field redefinition and, therefore, it is integrable.

We can also obtain a scalar Lax operator for the sNLS from the
Lax operator for the sTB system [12]. Using the field identifications
(5.17) in (3.9) we get
$$
\eqalign{
L=&D^2+{(D^3Q)\over(DQ)}-{\overline Q}Q-D^{-1}{\overline Q}(DQ)\cr
\noalign{\vskip 4pt}%
=&G{\widetilde L}G^{-1}
}\eqno(5.22)
$$
where
$$
\eqalign{
G=&(DQ)^{-1}\cr
{\widetilde L}=&D^2-{\overline Q}Q-(DQ)D^{-1}{\overline Q} \cr
}\eqno(5.23)
$$
We see that the two Lax operators, $L$ and ${\widetilde L}$, are related by a
gauge transformation in the superspace. Although, this resembles the bosonic
case (see eqs. (2.27) and (2.28)), ${\widetilde L}$ does not lead to any
consistent equation in the standard or nonstandard representation of the Lax
equation. However, the formal adjoint of ${\widetilde L}$
$$
{\cal L}={\widetilde L}^*=-\left(D^2+{\overline Q}Q-{\overline Q}
D^{-1}(DQ)\right)\eqno(5.24)
$$
gives the sNLS equations (5.19) via the nonstandard Lax equation
$$
{\partial{\cal L}\over\partial t}=\left[{\cal L},
\left({\cal L}^2\right)_{\ge1}\right]
\eqno(5.25)
$$
This differs from the bosonic case (as we had pointed out earlier), since
it is only the formal adjoint of the gauge
transformed Lax operator which leads to the consistent equations. The
other difference from the bosonic case is the fact that the supersymmetric
generalization of NLS is obtained as a nonstandard Lax equation in superspace.

The relations in (5.17) are invertible and can be formally written as
$$
\eqalignno{
Q=&\left(D^{-1}\hbox{e}^{\left(D^{-2}\left(-(D\Phi_0)+
\Phi_1(L^{-1}\Phi_0)\right)\right)}\right)
&(5.26a)\cr
{\overline Q}=&-\Phi_1
\hbox{e}^{\left(-D^{-2}\left(-(D\Phi_0)+\Phi_1(L^{-1}\Phi_0)\right)\right)}
&(5.26b)
}
$$
and these ((5.17) and (5.26)) define the connecting relations between the sNLS
and the sTB equations.

The conserved charges for the sNLS equation can be obtained  from
$$
H_n={1\over n}\hbox{sTr}\,{\cal L}^n \eqno(5.27)
$$
The first few conserved charges have the form
$$
\eqalign{
H_1=&\int dz\, (DQ){\overline Q}\cr
H_2=&\int dz\, (D^3Q){\overline Q}\cr
H_3=&\int dz\,\Bigl[(D^2{\overline Q})(DQ){\overline Q}Q-
(D^3Q)(D^2{\overline Q})-(D{\overline Q})(DQ)^2{\overline Q}\Bigr]\cr
}\eqno(5.28)
$$

Since the sTB and the sNLS are related through a field redefinition and we
already know the Hamiltonian structures of the sTB system, we can also obtain
the Hamiltonian structures of the sNLS equations [11] in the following way.
Let us define the transformation matrix between the variables of the two
systems as
$$
P = \left[{\delta(\Phi_0,\Phi_1)\over\delta(Q,{\overline Q})}\right]\eqno(5.29)
$$
where $\left[{\delta(\Phi_0,\Phi_1)\over\delta(Q,{\overline Q})}\right]$ is the
matrix formed from the Fr\'echet derivatives of $\Phi_0$ and $\Phi_1$ with
respect to $Q$ and ${\overline Q}$. Then, the Hamiltonian
structures of the two systems are related by [43]
$$
{\cal D}= P\,{\cal D}^{\,\hbox{\pq sNLS}} P^* \eqno(5.30)
$$
where $P^*$ is the formal adjoint of $P$ (with the matrix transposed).
Explicitly, we have
$$
P=\pmatrix{-D(DQ)^{-1}D+D^{-1}{\overline Q} & -D^{-1}Q\cr
\noalign{\vskip 10pt}%
-{\overline Q}D & -(DQ)}\eqno(5.31)
$$
It is easy to see that the matrix $P$  factorizes in the following form
$$
P={\widetilde P}G\eqno(5.32)
$$
where
$$
\eqalignno{
{\widetilde P}=&\pmatrix{-D^{-1}&-D^{-1}(DQ)^{-1}Q D^2\cr
\noalign{\vskip 10pt}%
0&-D^2}&(5.33a)\cr
\noalign{\vskip 15pt}%
G=&\pmatrix{-{\cal L}(DQ)^{-1}D&0\cr
\noalign{\vskip 10pt}%
D^{-2}{\overline Q}D &D^{-2}(DQ)} &(5.33b)
}
$$
This may suggest that there is another equation to which both the sTB and
the sNLS equations can be transformed [23]. But we will not go into a
discussion of this.

{}From  (5.30) we note that the Hamiltonian structures of the sNLS system
can be obtained from those of the sTB system (written in terms of $Q$ and
${\overline Q}$) as
$$
{\cal D}^{\,\hbox{\pq sNLS}}= P^{-1}{\cal D}\,(P^*)^{-1}
=G^{-1}{\widetilde P}^{-1}{\cal D}\,({\widetilde P}^*)^{-1}G^{-1}\eqno(5.34)
$$
where the inverse matrix has the form
$$
\eqalign{
P^{-1}=&
\pmatrix{D^{-1}(DQ){\cal L}^{-1}D&
-D^{-1}(DQ){\cal L}^{-1}Q(DQ)^{-1}\cr
\noalign{\vskip 10pt}%
-{\overline Q}{\cal L}^{-1}D &
-(1-{\overline Q}{\cal L}^{-1}Q)(DQ)^{-1} }\cr
\noalign{\vskip 12pt}%
=& G^{-1}{\widetilde P}^{-1}}\eqno(5.35)
$$
For completeness, we record here the form of the inverse matrices
$$
\eqalignno{
{\widetilde P}^{-1}=&\pmatrix{-D&(DQ)^{-1}Q \cr
\noalign{\vskip 10pt}%
0&-D^2}&(5.36a)\cr
\noalign{\vskip 15pt}%
G^{-1}=&\pmatrix{-D^{-1}(DQ){\cal L}^{-1}&0\cr
\noalign{\vskip 10pt}%
{\overline Q}{\cal L}^{-1} &(DQ)^{-1}D^2} &(5.36b)
}
$$

Now, using the second Hamiltonian structure (4.6) of the sTB system,
we obtain from (5.34)
$$
{\cal D}_2^{\,\hbox{\pq sNLS}} = P^{-1}{\cal D}_2(P^*)^{-1} \eqno(5.37)
$$
which alongwith (5.35) gives after some tedious algebra
$$
{\cal D}_2^{\,\hbox{\pq sNLS}}=
\pmatrix{-QD^{-1}Q&-{1\over2}D+QD^{-1}{\overline Q}\cr
\noalign{\vskip 10pt}%
-{1\over2}D+{\overline Q}D^{-1}Q&
-{\overline Q}D^{-1}{\overline Q} }
\eqno(5.38)
$$
This is second Hamiltonian structure that was derived in
ref. [42]. We note that it is enormously simpler to check this result by
noting that ${\cal D}_2$ factorizes as
$$
{\cal D}_2 = P\,{\cal D}_2^{\,\hbox{\pq sNLS}}P^* \eqno(5.39)
$$
with ${\cal D}_2^{\,\hbox{\pq sNLS}}$ given by (5.38).

The first Hamiltonian structure for the sNLS [11] can also be derived from
(5.34) as
$$
{\cal D}_1^{\,\hbox{\pq sNLS}} = P^{-1}{\cal D}_1(P^*)^{-1} \eqno(5.40)
$$
which upon using equations (4.3) and (5.35), gives
$$
{\cal D}_1^{\,\hbox{\pq sNLS}}=
\pmatrix{
-D^{-1}(DQ)\Delta(DQ)D^{-1}
&
D^{-1}(DQ)\Delta{\overline Q}\cr
&
+D^{-1}(DQ){\cal L}^{-1}D^2(DQ)^{-1}\cr
\noalign{\vskip 20pt}%
{\overline Q}\Delta(DQ)D^{-1}
&
-{\overline Q}{\cal L}^{-1}D^2(DQ)^{-1}-
(DQ)^{-1}D^2({\cal L}^*)^{-1}{\overline Q}\cr
+(DQ)^{-1}D^2({\cal L}^*)^{-1}(DQ)D^{-1}
&
-{\overline Q}\Delta{\overline Q}
}\eqno(5.41)
$$
where  we have defined
$$
\Delta\equiv
{\cal L}^{-1}\left(D^2((DQ)^{-1}Q)\right)({\cal L}^*)^{-1}
\eqno(5.42)
$$
Like the first Hamiltonian structure (5.16) of the sKdV equation, we note that
this structure is highly nonlocal.

It can now be checked explicitly that the sNLS equations (5.19) can be
written in the Hamiltonian form
$$
\partial_t\pmatrix{Q\cr
\noalign{\vskip 10pt}%
{\overline Q}}
={\cal D}_1^{\,\hbox{\pq sNLS}}
\pmatrix{{\delta H_{3}\over\delta Q}\cr
\noalign{\vskip 10pt}%
{\delta H_{3}\over\delta{\overline Q}}}=
{\cal D}_2^{\,\hbox{\pq sNLS}}
\pmatrix{{\delta H_{2}\over\delta Q}\cr
\noalign{\vskip 10pt}%
{\delta H_{2}\over\delta{\overline Q}}
}
\eqno(5.43)
$$
where the Hamiltonians are defined as in (5.28). A direct check of (5.43)
for the first Hamiltonian
structure (5.41) is quite involved. But we note that  it is much easier,
instead, to check that
$$
\pmatrix{{\delta H_{3}\over\delta Q}\cr
\noalign{\vskip 10pt}%
{\delta H_{3}\over\delta{\overline Q}}}=
\left({\cal D}_1^{\,\hbox{\pq sNLS}}\right)^{-1}
\pmatrix{ -(D^4 Q) + 2\left(D((DQ){\overline Q})\right)Q\cr
\noalign{\vskip 10pt}%
(D^4 \overline Q ) - 2\left(D((D{\overline Q})Q)\right){\overline Q}
}
\eqno(5.44)
$$
The two Hamiltonian structures (5.38) and (5.41) automatically
satisfy  the Jacobi
identity since the Hamiltonian structures of the sTB system do and define a
recursion operator which would relate all the Hamiltonian structures as well as
the conserved quantities of the system in a standard manner. However, as is
clear from the structure of the first Hamiltonian structure, this
recursion operator is extremely nontrivial and nonlocal.

\bigskip
\noindent {\bf 5.3 {Reduction to  Supersymmetric mKdV}}
\medskip

Finally, let us note that if we identify
$$
{\overline Q}=Q\eqno(5.45)
$$
in (5.24), then, the nonstandard Lax equation
$$
{\partial{\cal L}\over\partial t}=\left[{\cal L},
\left({\cal L}^3\right)_{\ge1}\right]
\eqno(5.46)
$$
yields
$$
{\partial  Q \over \partial  t} =-(D^6 Q) +3\left(D^2(Q(DQ))\right)(DQ)
\eqno(5.47)
$$
This is the supersymmetric mKdV equation [8] and shows that we can embed
the smKdV
equation in the sNLS much like the sKdV can be embedded in the sTB
equation [10]. However, as we have pointed out earlier, unlike the bosonic
case, it is the adjoint operator in equation (5.24) which leads to the
consistent dynamical equation. The Hamiltonian structures for smKdV can also be
derived along the lines of our earlier discussions. We leave the derivation to
the reader.

\bigskip
\noindent {\bf 6. {Nonstandard Supersymmetric KP Hierarchy}}
\medskip

As we saw in sec. 5.2 the sNLS equation can be obtained from the
nonstandard Lax equation (5.25) with the Lax operator (5.24).
If we rewrite the operator ${\cal L}$ in (5.24)
using the supersymmetric Liebnitz rule given in the Appendix
we get
$$
\eqalign{
{\cal L}=&-\left(D^2+{\overline Q}Q-{\overline Q}D^{-1}(DQ)\right)\cr
=&-\left(D^2+\sum_{n=-1}^\infty\Psi_n D^{-n}\right)\cr
}\eqno(6.1)
$$
where
$$
\eqalign{
\Psi_{-1}=&0\cr
\Psi_n=&(-1)^{[{n+1\over2}]}\,{\overline Q}(D^nQ),\quad n\ge 0}
\eqno(6.2)
$$
$\Psi_{2n}$ ($\Psi_{2n+1}$) are bosonic (fermionic)  superfields which
define constrained variables much like the bosonic case
(see (2.30) and (2.31)). The form of
$\cal L$ in (6.1) resembles the Lax operator for the sKP hierarchy
and, therefore, we can think of the sNLS system as a constrained sKP
system of nonstandard kind (Viewed in this way the sTB system, given by the Lax
operator (3.9), can also be thought of as a constrained,
nonstandard sKP system.)
However, we note that the Lax operator in the present case is an even parity
operator [39,44,45] and not of the usual Manin-Radul form [7]. Our result also
differs from the ones obtained in [14] through coset reduction.

This is a new system and, therefore, deserves further study [12]. Let us
consider a general supersymmetric Lax operator of the form (6.1)
$$
\eqalign{
L&=D^2+\Psi_0+\Psi_1D^{-1}+\Psi_2D^{-2}+\cdots\cr
 &=D^2+\sum_{n=0}^\infty\Psi_nD^{-n}
}\eqno(6.3)
$$
where the superfields $\Psi_n$ have the Grassmann parity
$$
|\Psi_n|={1-(-1)^n\over 2}\eqno(6.4)
$$
and have the form
$$
\eqalign{
 \Psi_{2n}=&q_{2n}+\theta \phi_{2n}\cr
 \Psi_{2n+1}=&\phi_{2n+1}+\theta q_{2n+1}\cr
}\eqno(6.5)
$$
with the $q_n$ ($\phi_n$) being the bosonic (fermionic) components of the
superfields.

The nonstandard flows associated with this sKP Lax operator are given by
$$
{\partial L\over\partial t_n}=\left[\left(L^n\right)_{\ge1},L\right]\eqno(6.6)
$$
For $n=1$, the flow is trivial and gives
$$
{\partial\Psi_n\over\partial t_1}=
(D^2\Psi_n)=\left({\partial\Psi_n\over\partial x}\right)\eqno(6.7)
$$
implying that the  time coordinate $t_1$ can be identified with $x$.

For $n=2$, the flow in  (6.6) gives
$$
\eqalign{
{\partial\Psi_n\over\partial t_2}=&
(D^4\Psi_n)+2(D^2\Psi_{n+2})+2\Psi_0(D^2\Psi_n)
+2\Psi_1(D\Psi_n)-2(1+(-1)^{n})\Psi_1\Psi_{n+1}\cr
&+2\sum_{\ell\ge1}\left\{-(-1)^{[{\ell/2}]}\left[\matrix{n+1\cr \ell}
\right]\Psi_{n-\ell+2}(D^{\ell}\Psi_0)
+(-1)^{[{\ell/2}]+n}\left[\matrix{n\cr \ell}\right]
\Psi_{n-\ell+1}(D^{\ell}\Psi_1)\right\}
}\eqno(6.8)
$$
where the super binomial coefficients $\left[\matrix{n\cr \ell}\right]$  are
defined in the Appendix. Equations for the bosonic components, $q$'s,
can be obtained from (6.8) [12]. In the bosonic limit -- when all
the $\phi_n$'s are zero -- if we set

$$
q_{2n}=0\quad\hbox{and}\quad q_{2n+1}=u_n\,,\quad\hbox{for all }n\eqno(6.9)
$$
we end up with
$$
\eqalign{
{\partial u_0\over\partial t_2}=&u_0''+2u_1'\cr
{\partial u_1\over\partial t_2}=&u_1''+2u_2'+2u_0u_0'\cr
{\partial u_2\over\partial t_2}=&u_2''+2u_3'-2u_0u_0''+4u_1u_0'\cr
\vdots&\cr
}\eqno(6.10)
$$
which are nothing other than the $t_2$-flows for the standard KP hierarchy
[32,46].

On the other hand, in the bosonic limit, if we set
$$
q_{2n+1}=0\quad\hbox{and}\quad q_{2n}=u_n\,,\quad\hbox{for all }n\eqno(6.11)
$$
then we obtain
$$
\eqalign{
{\partial u_0\over\partial t_2}=&u_0''+2u_1'+2u_0u_0'\cr
{\partial u_1\over\partial t_2}=&u_1''+2u_2'+2u_0u_1'+2u_0'u_1\cr
{\partial u_2\over\partial t_2}=&u_2''+2u_3'-2u_1u_0''+2u_0u_2'+4u_2u_0'\cr
\vdots&\cr
}\eqno(6.12)
$$
which are nothing other than the $t_2$-flows associated with the standard mKP
hierarchy [32,46].

For $n=3$, equation (6.6) gives
$$
\eqalign{
{\partial\Psi_n\over\partial t_3}=&
(D^6\Psi_n)+3(D^4\Psi_{n+2})+3(D^2\Psi_{n+4})
+3\Psi_0(D^4\Psi_n)+6\Psi_0(D^2\Psi_{n+2})\cr
&+3\Psi_1(D^3\Psi_n)+3\Psi_1(D\Psi_{n+2})-3(-1)^n\Psi_1(D^2\Psi_{n+1})\cr
&-3(1+(-1)^n)\Psi_1\Psi_{n+3}-
3(1+(-1)^n)((D^2\Psi_1)+\Psi_3+2\Psi_1\Psi_0)\Psi_{n+1}\cr
&+3((D^2\Psi_0)+\Psi_2+\Psi_0^2)(D^2\Psi_n)
+3((D^2\Psi_1)+\Psi_3+2\Psi_1\Psi_0)(D\Psi_n)\cr
&+3\sum_{\ell\ge1}\Biggl\{-(-1)^{[\ell/2]}
\left[\matrix{n+3\cr \ell}\right]\Psi_{n-\ell+4}(D^{\ell}\Psi_0)\cr
\noalign{\vskip -2pt}%
&\phantom{+3\sum_{\ell\ge1}\Biggl\{}+(-1)^{[{\ell/2}]+n}
\left[\matrix{n+2\cr \ell}\right]\Psi_{n-\ell+3}(D^{\ell}\Psi_1)\cr
\noalign{\vskip -2pt}%
&\phantom{+3\sum_{\ell\ge1}\Biggl\{}+(-1)^{[{\ell/2}]}
\left[\matrix{n+1\cr \ell}\right]\Psi_{n-\ell+2}(D^{\ell}
\left((D^2\Psi_0)+\Psi_2+\Psi_0^2)\right)\cr
\noalign{\vskip -2pt}%
&\phantom{+3\sum_{\ell\ge1}\Biggl\{}+(-1)^{[{\ell/2}]+n}
\left[\matrix{n\cr \ell}\right]\Psi_{n-\ell+1}(D^{\ell}
\left((D^2\Psi_1)+\Psi_3+2\Psi_1\Psi_0)\right)\Biggl.\Biggr\}
}\eqno(6.13)
$$
Bosonic components equations for $q$'s can again be obtained from  (6.13)
and in the bosonic limit, with the identifications in (6.9), we obtain
$$
\eqalign{
{\partial u_0\over\partial t_3}=&u_0'''+3u_1''+3u_2'+6u_0u_0'\cr
{\partial u_1\over\partial t_3}=&u_1'''+3u_2''+3u_3'+6u_0u_1'+6u_0'u_1\cr
\vdots&\cr
}\eqno(6.14)
$$
which are the $t_3$-flows for the standard KP hierarchy.
On the other hand, the identifications in (6.11) leads to
$$
\eqalign{
{\partial u_0\over\partial t_3}=&u_0'''+3u_1''+3u_2'+3u_0u_0''+3(u_0')^2+
6(u_1u_0)'+3u_0^2u_0'\cr
\vdots&\cr
}\eqno(6.15)
$$
These are  the $t_3$-flows for the standard mKP hierarchy. So, this new system,
namely, the nonstandard sKP hierarchy contains both the standard KP and the
standard mKP flows in its bosonic limit. The supersymmetry and the
nonstandard nature of the equation has unified bosonic hierarchies into a more
general one.

Since the  nonstandard sKP equation of (6.6) reduces to the standard KP flows
in the bosonic limit let us examine the nature of the sKP equation that
it leads to [12]. If we put
$$
\Psi_{2n}=0\,,\quad\hbox{for all }n\eqno(6.16)
$$
then, the first two nontrivial equations following from  (6.8) are
$$
\eqalign{
{\partial\Psi_1\over\partial t_2}=&(D^4\Psi_1)+2(D^2\Psi_3)\cr
\noalign{\vskip 4pt}%
{\partial\Psi_3\over\partial t_2}=&(D^4\Psi_3)+2(D^2\Psi_5)-2(D(\Psi_1\Psi_3))
+2\Psi_1(D^3\Psi_1)\cr
}\eqno(6.17)
$$
and the first nontrivial equation following from (6.13) has the form
$$
{\partial\Psi_1\over\partial t_3}=(D^6\Psi_1)+3(D^4\Psi_3)+3(D^2\Psi_5)
+3(D^2(\Psi_1(D\Psi_1)))\eqno(6.18)
$$
{}From  (6.17) and (6.18), we obtain, after the identifications
$t_2=y$, $t_3=t$ and $\Psi_1=\Psi=\phi+\theta u$, the equation
$$
D^2\left({\partial\Psi\over\partial t}-{1\over 4}(D^6\Psi)-
{3\over2}(D^2(\Psi(D\Psi)))-
{3\over2}(D(\Psi(\partial^{-1}{\partial\Psi\over\partial y})))\right)
={3\over 4}{\partial^2\Psi\over\partial y^2}\eqno(6.19)
$$
This is a  supersymmetric generalization of the KP equation which
differs from the usual Manin-Radul equation [7,47]
because of the presence of the nonlocal terms. In components this
equation yields
$$
\eqalign{
{\partial\ \over\partial x}\left({\partial u\over\partial t}-{1\over4}u'''
-3uu'+{3\over2}\phi\phi''
-{3\over2}\phi'(\partial^{-1}{\partial\phi\over\partial y})
-{3\over2}\phi{\partial\phi\over\partial y}\right)
=&{3\over4}{\partial^2 u\over\partial y^2}\cr
\noalign{\vskip 5pt}%
{\partial\ \over\partial x}\left({\partial\phi\over\partial t}
-{1\over4}\phi'''-{3\over2}(u\phi)'
-{3\over2}u(\partial^{-1}{\partial\phi\over\partial y})+
{3\over2}\phi(\partial^{-1}{\partial u\over\partial y})
\right)
=&{3\over4}{\partial^2 \phi\over\partial y^2}\cr
}\eqno(6.20)
$$
which in the bosonic limit, $\phi=0$, reduces to the KP equation. However,
these equations are not invariant under $y\leftrightarrow-y$
unlike the Manin-Radul equations. We also note that when we restrict
the variables $u$ and $\phi$  to be independent of $y$, these
equations reduce to the supersymmetric KdV equation [7,8].
Therefore, equation (6.19) represents  a new supersymmetric
generalization of the KP equation.

In the next section, we will show that the sTB hierarchy contains, in
addition to the conserved local charges, fermionic nonlocal charges that
are conserved and lead to higher symmetries.
\bigskip
\noindent {\bf 7. {Nonlocal Charges}}
\medskip

Occurrence of nonlocal conserved charges in supersymmetric integrable models
such as the sKdV equation was first recognized in [48]. In ref. [49], a
systematic procedure for their construction was given within the framework
of the Gelfand-Dikii formalism. It was shown
that while the local charges for the sKdV can
be obtained from odd powers of the  square root of the Lax operator,
$L^{2n-1\over2}$, the nonlocal ones
arise from odd powers of the quartic roots, $L^{2n-1\over4}$.
For the sTB, since the local charges $Q_n$'s are
related to integer powers of $L$ (eq. (3.13)), we can expect conserved
nonlocal charges
$F_n$'s  from odd powers of the square root of $L$ [13], that is,
$$
F_{2n-1\over2}=\hbox{sTr}\,L^{2n-1\over2}\qquad n=1,2,\dots\eqno(7.1)
$$

The square root of the Lax operator, $L$, for sTB (given in (3.9))  has
the form
$$
L^{1/2}=D+a_0+a_1D^{-1}+a_2D^{-2}+a_3D^{-3}+a_4D^{-4}+a_5D^{-5}+\cdots
\eqno(7.2)
$$
where
$$
\eqalign{
a_0=&2(D^{-2}\Phi_1)-\Phi_0\cr
a_1=&-(D^{-1}\Phi_1)\cr
a_2=&(D^{-1}(\Phi_0\Phi_1))-2(D^{-2}((D\Phi_0)\Phi_1))
+\Phi_0(D^{-1}\Phi_1)-\Phi_1\cr
a_3=&{1\over2}(D^{-1}\Phi_1)^2-(D\Phi_0)(D^{-1}\Phi_1)+
(D^{-1}((D\Phi_0)\Phi_1))+(D\Phi_1)\cr
\int dz\,a_5=&\int dz\,\biggl[{\cal O}
(Da_1)-\Bigl(D^{-1}\bigl(2a_3(D^2a_0)-a_1a_3+(Da_1)(D{\cal O})\bigr)\Bigr)
\biggr]}\eqno(7.3)
$$
and we have defined
$$
{\cal O}\equiv (D^2a_0)+(Da_1)-2a_2\eqno(7.4)
$$
with the grading of the coefficients given by
$$
|a_n|=n+1\eqno(7.5)
$$
The first three nonlocal charges can be obtained [13] after some tedious, but
straightforward, calculations to be
$$
\eqalign{
F_{1/2}=&-\int dz\, (D^{-1}\Phi_1)\cr
F_{3/2}=&-\int dz\,\biggl[{3\over2}(D^{-1}\Phi_1)^2-\Phi_0\Phi_1-
\Bigl(D^{-1}\bigl((D\Phi_0)\Phi_1\bigl)\Bigr)\biggr]\cr
F_{5/2}=&-\int dz\,\biggl[\,{1\over6}(D^{-1}\Phi_1)^3-
\bigl(5(D^{-2}\Phi_1)\Phi_1
-2\Phi_0\Phi_1-3(D\Phi_1)-(D^{-1}\Phi_1)^2\bigr)(D\Phi_0)\cr
&\phantom{2\int dz\,\Bigl[}
+\Bigl(D^{-1}\bigl((D\Phi_1)\Phi_1+\Phi_1(D\Phi_0)^2-
(D\Phi_1)(D^2\Phi_0)\bigr)\Bigr)\biggr]\cr
}\eqno(7.6)
$$
These charges have been explicitly checked to be conserved under the flow
(3.10). (Conservation, of course, follows from the structure of the Lax
equation (3.7). But an explicit check assures that the form of the
charges given in (7.6) are indeed correct.)

Let us note that all the nonlocal charges in (7.6) are
fermionic and $[F_{2n-1\over2}]={2n-1\over2}$. Also, even though these
charges are expressed as superspace integrals, they are not invariant under the
supersymmetry transformations (3.12). This is
because while the superspace integral of a local function of superfields is
automatically supersymmetric, this is not necessarily true for nonlocal
functions. However, this is not a matter of worry since even the supersymmetry
charge, in these integrable models, is not supersymmetric.

We note that the nonlocal charges of the sTB hierarchy in (7.6) reduce to
those of the sKdV hierarchy, up to normalizations [49],
when we set $\Phi_0=0$. This is not surprising
since we have already shown that the sKdV system is contained in the
sTB system. However, unlike the sKdV system, the nonlocal charges (7.6) are
not related recursively by either $R$ or $R^\dagger$, given by (4.9) and (4.20)
respectively. It is likely that these fermionic
charges  generate fermionic flows with distinct Hamiltonian
structures of their own
which in turn can give a ``recursion'' operator connecting them.
In [50] odd flows based on Jacobian
sKP hierarchies were studied for the sKdV case, and maybe this result can be
generalized to the sTB hierarchy if we use, instead, the nonstandard
sKP hierarchy of sec. 6. We also note here that using the
transformation (5.17)  we can obtain the nonlocal charges for the sNLS
equation which should coincide with the ones constructed in [41].

Let us note that the
supersymmetry transformations (3.12) are generated by the
conserved fermionic charge
$$
Q=-\int dx\,\left(\psi_1J_0+\psi_0J_1\right)\eqno(7.7)
$$
through the first Hamiltonian structure in (4.5) as
$$
\eqalign{
\delta_Q J_0=&\epsilon\{Q,J_0\}_1=\epsilon\psi_0'\cr
\delta_Q J_1=&\epsilon\{Q,J_1\}_1=\epsilon\psi_1'\cr
\delta_Q\psi_0=&\epsilon\{Q,\psi_0\}_1=\epsilon J_0\cr
\delta_Q\psi_1=&\epsilon\{Q,\psi_1\}_1=\epsilon J_1\cr
}\eqno(7.8)
$$

We can also show easily with eqs. (7.7) and (4.5) that
$$
\{Q,Q\}_1=-Q_2\eqno(7.9)
$$
which implies that the supersymmetry charge is not supersymmetric --
rather it satisfies a graded Lie algebra.
As we have mentioned earlier, the nonlocal charges are also not invariant
under supersymmetry transformations (3.12) or (7.8) and, in fact, we obtain
$$
\eqalign{
\{Q,F_{1/2}\}_1=&Q_1 \cr
\{Q,F_{3/2}\}_1=&{1\over2}Q_2 \cr
\{Q,F_{5/2}\}_1=&{1\over3}Q_3+{1\over24}Q_1^3 \cr
}\eqno(7.10)
$$
Furthermore, since the supersymmetry  charge $Q$ as well as the nonlocal
charges $F_{2n-1\over2}$ are conserved, they are in involution with all
the local conserved charges $Q_n$, i.e,
$$
\eqalign{
\{Q_n,Q_m\}_1=&0\cr
\{Q_n,F_{2m-1\over2}\}_1=&0\cr
\{Q_n,Q\}_1=&0\cr
}\eqno(7.11)
$$

The algebra of the  nonlocal charges is quite interesting as well and we
list here only the first few relations
$$
\displaylines{
\hfill
\eqalign{
\{F_{1/2},F_{1/2}\}_1=&0\cr
\{F_{1/2},F_{3/2}\}_1=&Q_1 \cr
\{F_{1/2},F_{5/2}\}_1=&Q_2 \cr
}
\hfill
\eqalign{
\{F_{3/2},F_{3/2}\}_1=&2Q_2 \cr
\{F_{3/2},F_{5/2}\}_1=&{7\over3}Q_3+{7\over24}Q_1^3 \cr
\{F_{5/2},F_{5/2}\}_1=&3Q_4-{5\over 8}Q_1^2 Q_2 \cr
}
\hfill
(7.12)}
$$
This  shows that the algebra of the
conserved charges $Q$, $Q_n$ and $F_{2n-1\over2}$, at least at these orders,
closes with respect to the
first Hamiltonian structure in (4.5). However, the algebra is not a linear Lie
algebra, but it is a graded nonlinear algebra where the nonlinearity
manifests in a cubic term. The canonical dimensions of the charges
allows for higher order nonlinearity to be present even in the algebraic
relations of these lower order charges, but
the algebra appears to involve only a cubic nonlinearity.
The Jacobi identity is seen to be trivially satisfied for this algebra since
the $Q_n$'s are in involution with all the fermionic charges ($Q_n$'s are the
Hamiltonians for the system and the fermionic charges are conserved.).
We note here that the cubic terms in (7.10) and (7.12)
arise from boundary contributions when nonlocal terms are involved. This can be
illustrated using the following realization of the inverse  operator
$$
\bigl(\partial^{-1} f(x)\bigr) ={1\over 2}\int d y \, \epsilon(x-y)f(y)
\,,\qquad \epsilon(x)= \cases{-1,\,& $x<0$\cr \phantom{-}0,\, &$x=0$\cr
+1,\, &$x>0$\cr}
\quad \eqno(7.13)
$$
This will lead, for instance, to boundary terms of the type
$$
\int dz\,(D^{-1}\Phi_1)^2\Phi_1={1\over3}\int dz\,D(D^{-1}\Phi_1)^3=
-{1\over12}Q_1^3\eqno(7.14)
$$
and this is the origin of the nonlinear terms.

Since the sTB hierarchy has three distinct Hamiltonian structures,
one can ask  other interesting questions such as what transformations does $Q$
generate with the second structure or what is the algebra of the charges with
respect to the second structure and so on.
In fact, it has been shown in [13] that the
charges do satisfy a graded algebra
with respect to the second structure as well and the algebra continues to be
a cubic algebra. In fact, the graded algebra obtained with the second
structure appears to be sort of a shifted version of the previous one.

{}From (7.12) we see that the nonlocal charge
$F_{3/2}$ has the same Poisson bracket with itself (except for normalizations)
as the supersymmetry charge (see (7.9)). Thus, we can think of $F_{3/2}$ as
also generating a set of supersymmetry transformations given by
$$
\eqalign{
\delta_{3/2} J_0=&\epsilon\{F_{3/2},J_0\}_1=\epsilon(\psi_0'-3\psi_1)\cr
\delta_{3/2} J_1=&\epsilon\{F_{3/2},J_1\}_1=-2\epsilon\psi_1'\cr
\delta_{3/2}\psi_0=&\epsilon\{F_{3/2},\psi_0\}_1=\epsilon J_1\cr
\delta_{3/2}\psi_1=&\epsilon\{F_{3/2},\psi_1\}_1=
\epsilon\left(3(\partial^{-1}J_1)-2J_0\right)\cr
}\eqno(7.15)
$$
These are distinct from the transformations in (3.12) and are nonlocal.
Moreover, $F_{3/2}$ can be seen from (7.12) and (7.10) to have the same
algebra, with the other generators of the algebra, as the supersymmetry charge,
$Q$. This
suggests that even though the sTB equation we have constructed, in section 3,
appears to have an $N=1$ supersymmetry, in reality, it has an $N=2$ extended
supersymmetry [15,16]. We will return to this question in some more detail
in the sec. 8.

For the present, let us note that the graded algebra of our system appears to
be a cubic algebra.
Cubic algebras have been found earlier in studies of other systems such as the
Heisenberg spin chains and the nonlinear sigma models [51-54]
and appear to be a
common feature when nonlocal charges are involved.
The nonlinearity in these algebras can, in principle, be high. However, the
interesting thing about these algebras is that it
is possible to redefine the generators in a highly nonlinear and nontrivial way
such that the algebra becomes a  cubic algebra. This is quite well known in
the case of the nonlinear sigma model [54] and we will describe below how this
is achieved in the case of supersymmetric integrable systems through the
example of sKdV (because it is a simpler system). We note here that
cubic terms are also present in the algebra of nonlocal charges,
in the case of the sKdV system [13], even though it has not been observed
before. If we take the sKdV equation,
$\Phi_t=-(D^6\Phi)+3D^2(\Phi(D\Phi))$ and use the nonlocal charges, the Lax
operator $L$
as well as the second Hamiltonian structure given in [49] (This differs from
(5.12) by constant factors and field redefinitions.),
and take the conserved local charges of sKdV as
$$
H_{2n-1}={2^{2n-1}\over 2n-1}\hbox{Tr}\,L^{2n-1\over2}\eqno(7.16)
$$
the following algebra can be obtained in a straightforward manner
$$
\displaylines{
\hfill
\eqalign{
\{J_{1/2},J_{1/2}\}=&-H_1\cr
\{J_{1/2},J_{3/2}\}=&0\cr
\{J_{1/2},J_{5/2}\}=&-6H_3-{1\over4}H_1^3\cr
}
\hfill
\eqalign{
\{J_{3/2},J_{3/2}\}=&4H_3-{1\over3}H_1^3\cr
\{J_{3/2},J_{5/2}\}=&0\cr
\{J_{5/2},J_{5/2}\}=&-36H_5-{9\over80}H_1^5\cr
}
\hfill(7.17)
}
$$
The important point to note
is the appearance of the quintic term in the bracket of $J_{5/2}$ with itself.
However, we can redefine
$$
\eqalign{
{\hat J}_{1/2}=&J_{1/2}\cr
{\hat J}_{3/2}=&J_{3/2}\cr
{\hat J}_{5/2}=&J_{5/2}+\alpha H_1^2 J_{1/2}\cr
}\eqno(7.18)
$$
where $\alpha$ can be chosen such that the algebra becomes cubic. From this
we strongly believe that one can redefine the charges even in
the case of sTB such that the right hand side of the algebra in (7.10) and
(7.12) will have the closed form structure
$$
a\,{\hat Q}_n+b\!\!\!\sum_{p+q+\ell=n}\!\!\!
{\hat Q}_p {\hat Q}_q {\hat Q}_\ell\eqno(7.19)
$$
where $a$ and $b$ are numerical factors and $n$ is the sum of the canonical
dimensions of the left hand side of the algebra.

We conclude this section by noting  that cubic algebras of this sort
can be  related to Yangians [51-55]. So, it
is natural to expect
that the algebra, in the present systems (both sKdV and sTB) also corresponds
to a Yangian. There is, however, a difficulty with this. Namely, a Yangian
starts out with a non-Abelian Lie algebra for the local charges. Here, in
contrast, the algebra of the local charges, $Q_n$'s, is involutive (Abelian).
There may still be an underlying Yangian structure in this algebra and this
remains an open question.

\bigskip
\noindent {\bf 8. {Conclusions}}
\medskip

We have discussed in this paper the supersymmetric generalization of the two
boson system and its various properties [10]. The supersymmetric
system, much like
its bosonic counterpart, appears to have a rich structure. It reduces to
various other known supersymmetric integrable systems. It has a bi-Hamiltonian
structure and since it reduces to various other supersymmetric
systems under field redefinitions or reductions,
allows us to construct various
quantities associated with these systems in a simple way. For example, as we
have pointed out, the Hamiltonian structures associated with various systems
can be obtained in a natural way [11]. The scalar
Lax operator associated with the
sNLS system follows quite simply and shows that it can be thought of as a
constrained, nonstandard sKP hierarchy [12]. The constrained, nonstandard
sKP hierarchy is in itself quite interesting because in addition to
giving a new integrable
supersymmetrization of the KP equation, it unifies all the KP and mKP flows
into a single equation. We have also derived the nonlocal charges associated
with the sTB system and their algebra appears to be a cubic algebra [13].
Most important, however, is the fact that even though we started with an $N=1$
supersymmetric system, the system appears to have developed an $N=2$ extended
supersymmetry. We now return to a discussion of this and other recent results
associated with this system [14-17].

We begin by noting that recently an $N=2$ supersymmetrization of the Two Boson
system has been obtained in ref. [15] from a direct
supersymmetrization of the second
Hamiltonian structure (2.18). The set of equations that result from [15]
is given by
$$
{\partial J\over \partial t}=[\,{\widetilde D},{\overline{\widetilde D}}\,]J'
+4J'J\eqno(8.1)
$$
where $J$ is the $N=2$ bosonic superfield (We have made the identifications
$S=J_0$ and $R=J_1$.)
$$
J={1\over 2}J_0+{\widetilde\theta}\xi+
{\overline{\widetilde\theta}}\,{\overline\xi}+
{1\over2}{\widetilde\theta}\,{\overline{\widetilde\theta}}
\left({1\over2}J'_0-J_1\right)\eqno(8.2)
$$
and ${\widetilde D}$ and ${\overline{\widetilde D}}$ are covariant
derivatives (We are using tilde variables so that
comparison with our superspace results in sec. 3 becomes simpler.)
$$
\eqalign{
{\widetilde D}=&{\partial\
\over\partial{\widetilde\theta}}-{1\over2}{\overline{\widetilde\theta}}
{\partial\ \over\partial x}\cr
\noalign{\vskip 4pt}%
{\overline{\widetilde D}}
=&{\partial\ \over\partial{\overline{\widetilde\theta}}}
-{1\over2}{\widetilde\theta}{\partial\ \over\partial x}\cr
}\eqno(8.3)
$$
which satisfy
$$
\{{\widetilde D},{\overline{\widetilde D}}\}=-{\partial\ \over\partial
x}\,,\qquad\{{\widetilde D},{\widetilde D}\}=
\{{\overline{\widetilde D}},{\overline{\widetilde D}}\}=0\eqno(8.4)
$$
In components, (8.1) takes the form
$$
\eqalign{
{\partial  J_0 \over \partial  t} &= -J''_0+2J'_0J_0+2J'_1\cr
\noalign{\vskip 4pt}%
{\partial  J_1 \over \partial  t} &= J''_1+2(J_1J_0)'+8(\xi{\overline\xi})'\cr
\noalign{\vskip 4pt}%
{\partial  {\overline\xi} \over \partial  t} &={\overline\xi}''
+2({\overline\xi} J_0)'\cr
\noalign{\vskip 4pt}%
{\partial  \xi \over \partial  t} &=\xi''+2(\xi J_0)'\cr
}
\eqno(8.5)
$$
It is now easy to see, with the linear identifications,
$$
\eqalign{
{\overline\xi}=&-{1\over2}(\psi_0'-\psi_1)\cr
\xi=&{1\over2}\psi_1\cr
}
\eqno(8.6)
$$
that the set of equations in (8.5) are nothing other than the sTB
equation  (3.11). As we have pointed out earlier,
the $N=2$ supersymmetry is already contained in our algebra (7.9), (7.10)
and (7.12). To compare the two results, we note that the $N=2$ transformations
of ref. [15] can be read out from the form of the superfield (8.2) and have
the following form in components
$$
\displaylines{
\hfill
\eqalign{
\delta J_0=&2\epsilon\xi\cr
\delta J_1=&2\epsilon\xi'\cr
\delta{\overline\xi}=&-{1\over2}\epsilon (J'_0-J_1)\cr
\delta\xi=&0\cr
}
\hfill
\eqalign{
{\overline\delta} J_0=&2\epsilon{\overline\xi}\cr
{\overline\delta} J_1=&0\cr
{\overline\delta}\,{\overline\xi}=&0\cr
{\overline\delta}\xi=&-{1\over2}\epsilon J_1\cr
}
\hfill(8.7)
}
$$
Using the identifications (8.6) we can write them in terms of the variables in
equation (3.11) as
$$
\displaylines{
\hfill
\eqalign{
\delta J_0=&\epsilon\psi_1\cr
\delta J_1=&\epsilon\psi_1'\cr
\delta\psi_0=&\epsilon\left(J_0-(\partial^{-1}J_1)\right)\cr
\delta\psi_1=&0\cr
}
\hfill
\eqalign{
{\overline\delta}J_0=&-\epsilon(\psi_0'-\psi_1)\cr
{\overline\delta}J_1=&0\cr
{\overline\delta}\psi_0=&-\epsilon(\partial^{-1}J_1)\cr
{\overline\delta}\psi_1=&-\epsilon J_1\cr
}
\hfill(8.8)
}
$$
{}From the supersymmetry transformations (7.8) and (7.15) we immediately
see that
$$
3\delta=\delta_Q-\delta_{3/2}\qquad\hbox{and}\qquad
-3{\overline\delta}=2\delta_Q+\delta_{3/2}\eqno(8.9)
$$
showing that the $N=2$ supersymmetry of ref. [15] is the
same $N=2$ supersymmetry
that is already present in our sTB system, but is manifest in the variables of
ref. [15].

We can translate our Hamiltonian structures to the new variables in a simple
way through the discussions of section 5. To obtain
the second Hamiltonian structure, ${\widetilde{\cal D}}_2$, in terms of the
variables in eq. (8.5) from the ones in eq. (4.7), we proceed as
in sec. 5.2.  Defining the matrix formed from the Fr\'echet derivatives of
$J_0,J_1,{\overline\xi},\xi$ with respect to $J_0,J_1,\psi_0,\psi_1$ as
$$
P = \left[{\delta(J_0,J_1,{\overline\xi},\xi)\over
\delta(J_0,J_1,\psi_0,\psi_1)}\right]\eqno(8.10)
$$
and relating the two Hamiltonian structures (see (5.30)) by
$$
{\widetilde{\cal D}}_2= P\,{\cal D}_2 P^* \eqno(8.11)
$$
where
$$
P=\pmatrix{
1 & 0 & 0 & 0\cr
\noalign{\vskip 3pt}%
0 & 1 & 0 & 0\cr
\noalign{\vskip 3pt}%
0 & 0 & -{1\over2}\partial & {1\over2}\cr
\noalign{\vskip 3pt}%
0 & 0 & 0 & {1\over2}
}\eqno(8.12)
$$
we obtain in a straightforward manner (${\cal D}_2$ is given in (4.7).)
$$
{\widetilde{\cal D}}_2=\pmatrix{
2\partial & \partial J_1-\partial^2 &-{\overline\xi} & \xi\cr
\noalign{\vskip 12pt}%
J_0\partial +\partial^2 & \partial J_1+J_1\partial &\xi\partial &
\partial\xi+\xi\partial\cr
\noalign{\vskip 12pt}%
{\overline\xi}&\partial{\overline\xi} & 0 &
-{1\over 4}(J_1-\partial J_0+\partial^2)\cr
\noalign{\vskip 12pt}%
-\xi & \partial\xi+\xi\partial &-{1\over 4}(J_1-\partial J_0+\partial^2) &0
}\eqno(8.13)
$$
Written in components, the only nonvanishing Poisson brackets are given by
$$
\eqalign{
\noalign{\vskip 3pt}%
\{J_0(x),J_0(y)\}_2=&2\delta'(x-y)\cr
\noalign{\vskip 3pt}%
\{J_0(x),J_1(y)\}_2=&(J_0\delta(x-y))'-\delta''(x-y)\cr
\noalign{\vskip 3pt}%
\{J_0(x),{\overline\xi}(y)\}_2=&-{\overline\xi}\delta(x-y)\cr
\noalign{\vskip 3pt}%
\{J_0(x),\xi(y)\}_2=&\xi\delta(x-y)\cr
\noalign{\vskip 3pt}%
\{J_1(x),J_1(y)\}_2=&J'_1\delta(x-y)+2J_1\delta'(x-y)\cr
\noalign{\vskip 3pt}%
\{J_1(x),{\overline\xi}(y)\}_2=&{\overline\xi}\delta'(x-y)\cr
\noalign{\vskip 3pt}%
\{J_1(x),\xi(y)\}_2=&\xi'\delta(x-y)+2\xi\delta'(x-y)\cr
\noalign{\vskip 3pt}%
\{{\overline\xi}(x),\xi(y)\}_2=&
-{1\over4}J_1\delta(x-y)+{1\over4}(J_0\delta(x-y))'-{1\over4}\delta''(x-y)\cr
}\eqno(8.14)
$$
It is worth noting here that the algebra, in terms of the new variables,
is local and that it is nothing other than the twisted $N=2$ superconformal
algebra [31] whose bosonic limit is the Virasoro-Kac-Moody algebra (2.18).

We also note that the first Hamiltonian structure can similarly be
transformed into the new variables using (8.11), (8.12) and (4.4) and
takes the form
$$
{\widetilde{\cal D}_1} = \pmatrix{0 & \partial & 0 & 0\cr
\noalign{\vskip 5pt}%
\partial & 0 & 0 & 0\cr
\noalign{\vskip 5pt}%
0 & 0 & 0 & {1\over 4}\partial\cr
\noalign{\vskip 5pt}%
0 & 0 & -{1\over 4}\partial & 0\cr}\eqno(8.15)
$$
which gives the following nonvanishing Poisson brackets in components
$$
\eqalign{
\{J_0(x),J_1(y)\}_1 =& \delta'(x-y)\cr
\noalign{\vskip 3pt}%
\{\xi(x),\xi(y)\}_1 =& {1\over 4}\delta'(x-y)}\eqno(8.16)
$$
This shows that both the first and the second Hamiltonian structures for
the system are local in the new variables and that the sTB system, in this
sense, is truly a bi-Hamiltonian system.

To compare the relation between the $N=2$ superfield of ref. [15] and our
superfields in eq. (3.3), we note the following. Given $J$ in (8.2), we
can define two chiral superfields which with the identification in
(8.6) take  the following form.
$$
\eqalign{2{\widetilde D}J =&\exp{\left({1\over2}{\widetilde\theta}\,
{\overline{\widetilde\theta}}\,{\partial\over\partial x}\right)} (\psi_1-
{\overline{\widetilde\theta}}J_1) = \exp{\left({1\over 2}{\widetilde\theta}\,
{\overline{\widetilde\theta}}\,{\partial\over\partial x}\right)}j\cr
2{\overline{\widetilde D}}{J} =& \exp{\left(-{1\over 2}
{\widetilde\theta}\,{\overline{\widetilde\theta}}\,
{\partial\over\partial x}\right)}
\left[(\psi_1-\psi'_0) + {\widetilde\theta}(J_1-J'_0)\right] =
\exp{\left(-{1\over 2}{\widetilde\theta}\,{\overline{\widetilde\theta}}\,
{\partial\over\partial x}\right)}{\overline j}}\eqno(8.17)
$$
It is also easy to see that we can define
$$
\eqalign{ \theta =&{1\over 2}({\widetilde\theta}-
{\overline{\widetilde\theta}})\cr
{\overline\theta} =&{1\over 2}({\widetilde\theta}+
{\overline{\widetilde\theta}})}\eqno(8.18)
$$
so that we can write the covariant derivative of section 3 (see eq. (3.1)) as
$$
D = {\widetilde D}-{\overline{\widetilde D}}\eqno(8.19)
$$
If we now set ${\overline\theta}=0$, then it is easy to see from (8.17)
that
$$
\eqalign{j =&\Phi_1\cr
{\overline j} =&(\Phi_1-\Phi'_0)}\eqno(8.20)
$$
In other words, the $N=1$ superfields are chirally related to the $N=2$
superfields under appropriate definition of coordinates.

To conclude, we note that there are still some very interesting questions, in
connection with the sTB system, that need further investigation. Among other
things, a derivation of the Hamiltonian structures from a generalization of the
Gelfand-Dikii brackets for supersymmetric nonstandard system remains an open
question. Similarly, further analysis of the structure of the algebra of
nonlocal charges as well as its possible connection with (super) Yangian is
quite interesting.

\bigskip
\noindent {\bf Acknowledgements}
\medskip

This work was supported in part by the U.S. Department of Energy Grant No.
DE-FG-02-91ER40685. J.C.B. would like to thank CNPq, Brazil, for
financial support.

\vfill\eject
\bigskip
\noindent {\bf {Appendix}}
\medskip

The supersymmetric Leibnitz [7] rule is given by
$$
D^kA=\sum_{j=0}^\infty \left[\matrix{k\cr j}\right] (-1)^{|A|(k+j)}
(D^j A)D^{k-j}\eqno(A.1)
$$
where for $k\ge0$
$$
\left[\matrix{k\cr j}\right]=
\cases{\pmatrix{[k/2]\cr
\noalign{\vskip 5pt}%
[j/2]}& for $k\ge j$ and
$(k,j)\not=(0,1)$ mod 2\cr
\noalign{\vskip 10pt}%
0 &otherwise\cr}\eqno(A.2)
$$
and for $k<0$
$$
\left[\matrix{k\cr j}\right]=(-1)^{[j/2]}\left[\matrix{-k+j-1\cr j}\right]
\eqno(A.3)
$$
where $[k/2]$ denotes the integral part of $k/2$ and $|A|$ is the Grassmann
parity of $A$, and $|A|=0\,(1)$ for $A$ even (odd).

The relation
$$
(-1)^{|A|}\bigl(D^{-1}(AB)\bigr)=A(D^{-1}B)-
\Bigl(D^{-1}\bigl((DA)(D^{-1}B)\bigr)\Bigr)\eqno(A.4)
$$
is very useful and can be easily proved through the Leibnitz rule. Also, to
perform integration by parts we need the generalized formula [35,49]
which holds for local functions $A$ and $B$,
$$
\int dz\,(D^nA)B=(-1)^{n|A|+{n(n+1)\over2}}\int dz\, A(D^nB),\quad\hbox{for
all }n\eqno(A.5)
$$
We note, however, that for nonlocal functions, the surface terms can not always
be neglected. Throughout this paper the parenthesis limit the action of the
inverse (integral) operators.

In the supersymmetric case the Poisson Brackets satisfy
$$
\eqalignno{
\{A,B\}=&-(-1)^{|A||B|}\{B,A\} &(A.6a)\cr
\{AB,C\}=&(-1)^{|B||C|}\{A,C\}B+A\{B,C\} &(A.6b)\cr
\{A,BC\}=&(-1)^{|A||B|}B\{A,C\}+\{A,B\}C &(A.6c)\cr
}
$$
and the Jacobi identity is
$$
(-1)^{|A||C|}\{\{A,B\},C\}+(-1)^{|B||C|}\{\{C,A\},B\}
+(-1)^{|A||B|}\{\{B,C\},A\}=0\eqno(A.7)
$$

\vfill\eject

\noindent {\bf {References}}
\bigskip

\item{1.} L.D. Faddeev and L.A. Takhtajan, ``Hamiltonian Methods in
the Theory of Solitons'' (Springer, Berlin, 1987).

\item{2.} A. Das, ``Integrable Models'' (World Scientific, Singapore,
1989).

\item{3.} L. A. Dickey, ``Soliton Equations and Hamiltonian Systems'' (World
Scientific, Singapore, 1991).

\item{4.} D. J. Gross and A. A. Midgal, Phys. Rev. Lett. {\bf 64}, 127 (1990);
D. J. Gross and A. A. Midgal, Nucl. Phys. {\bf B340}, 333 (1990); E. Br\'ezin
and V. A. Kazakov, Phys. Lett. {236B}, 144 (1990); M. Douglas and S. H.
Shenker,
Nucl. Phys. {\bf B335}, 635 (1990);
A. M. Polyakov in ``Fields, Strings and Critical Phenomena'', Les
Houches 1988, ed. E. Br\'ezin and J. Zinn-Justin (North-Holland, Amsterdam,
1989); L. Alvarez-Gaum\'e, Helv. Phys. Acta {\bf 64}, 361 (1991); P. Ginsparg
and G. Moore, ``Lectures on 2D String Theory and 2D Gravity'' (Cambridge, New
York, 1993).

\item{5.} L. Alvarez-Gaum\'e and J. L. Man\~es, Mod. Phys. Lett. {\bf A6},
2039 (1991); L. Alvarez-Gaum\'e, H. Itoyama, J. Man\~es and A. Zadra, Int. J.
Mod. Phys. {\bf A7}, 5337 (1992).

\item{6.} S. Stanciu, ``Supersymmetric Integrable Hierarchies and String
Theory'', Bonn University preprint, BONN-IR-94-07 (see also hep-th/9407189); M.
Becker, ``Non-Perturbative Approach to 2D-Supergravity and Super-Virasoro
Constraints'', CERN preprint, CERN-TH.7173/94 (see also hep-th/9403129); and
references therein.

\item{7.} Y. I. Manin and A. O. Radul, Commun. Math. Phys. {\bf 98}, 65 (1985).

\item{8.} P. Mathieu, J. Math. Phys. {\bf 29}, 2499 (1988).

\item{9.} B.A. Kupershmidt, Commun. Math. Phys. {\bf 99}, 51 (1985).

\item{10.} J. C. Brunelli and A. Das, Phys. Lett. {\bf B337}, 303 (1994).

\item{11.} J. C. Brunelli and A. Das, ``Bi-Hamiltonian Structure of the
Supersymmetric Nonlinear Schr\"odinger Equation'', University of Rochester
preprint UR-1421 (1995) (also hep-th/9505041).

\item{12.} J. C. Brunelli and A. Das, ``A Nonstandard Supersymmetric KP
Hierarchy'', University of Rochester preprint UR-1367 (1994) (also
hep-th/9408049), to appear in the Rev. Math. Phys..

\item{13.} J. C. Brunelli and A. Das, ``Properties of Nonlocal Charges in the
Supersymmetric Two Boson Hierarchy'', University of Rochester
preprint UR-1417 (1995) (also hep-th/9504030).

\item{14.} F. Toppan, Int. J. Mod. Phys. {\bf A10}, 895 (1995).

\item{15.} S. Krivonos and A. Sorin, ``The Minimal $N=2$ Superextension of the
NLS Equation'', preprint JINR E2-95-172 (also hep-th/9504084).

\item{16.} S. Krivonos, A. Sorin and F. Toppan, ``On the Super-NLS Equation and
its Relation with $N=2$ Super-KdV within Coset Approach'', preprint
JINR E2-95-185 (also hep-th/9504138).

\item{17.} Z. Popowicz, Phys. Lett. {\bf A194}, 375 (1994).

\item{18.} L.J.F. Broer, Appl. Sci. Res. {\bf 31}, 377 (1975).

\item{19.} D.J. Kaup, Progr. Theor. Phys. {\bf 54}, 396 (1975).

\item{20.} H. Aratyn, L.A. Ferreira, J.F. Gomes and A.H. Zimerman, Nucl.
Phys. {\bf B402}, 85 (1993); H. Aratyn, L.A. Ferreira, J.F. Gomes and A.H.
Zimerman, ``On $W_\infty$ Algebras, Gauge Equivalence of KP Hierarchies,
Two-Boson Realizations and their KdV Reductions'', in Lectures at the VII
J. A. Swieca Summer School, S\~ao Paulo, Brazil, January 1993,
eds. O. J. P. \'Eboli and V. O. Rivelles (World Scientific, Singapore, 1994);
H. Aratyn, E. Nissimov and S. Pacheva, Phys. Lett. {\bf B314}, 41 (1993).

\item{21.} L. Bonora and C.S. Xiong, Phys. Lett. {\bf B285}, 191 (1992); L.
Bonora and C.S. Xiong, Int. J. Mod. Phys. {\bf A8}, 2973 (1993).

\item{22.} M. Freeman and P. West, Phys. Lett. {\bf 295B}, 59 (1992).

\item{23.} J. Schiff, ``The Nonlinear Schr\"odinger Equation and
Conserved Quantities in the Deformed Parafermion and SL(2,{\bf R})/U(1)
Coset Models'', Princeton preprint IASSNS-HEP-92/57 (1992)
(also hep-th/9210029).

\item{24.} D. J. Benney, Stud. Appl. Math. {\bf L11}, 45 (1973).

\item{25.} J. C. Brunelli, A. Das and W.-J. Huang, Mod. Phys. Lett. {\bf 9A},
2147 (1994).

\item{26.} A. Das and W.-J. Huang, J. Math. Phys. {\bf 33}, 2487 (1992).

\item{27.} W. Oevel and W. Strampp, Commun. Math. Phys. {\bf 157}, 51 (1993).

\item{28.} A. Das and S. Roy, J. Math. Phys. {\bf 32}, 869 (1991).

\item{29.} W.J. Huang, J. Math. Phys. {\bf 35}, 993 (1994).

\item{30.} K. Bardakci and M. B. Halpern, Phys. Rev. {\bf D3}, 2493 (1971).

\item{31.} E. Witten, Commun. Math. Phys. {\bf 117}, 353 (1988); {\bf 118},
411 (1988); Nucl. Phys. {\bf B340} 281 (1990); T. Eguchi and S. Yang, Mod.
Phys.
Lett. {\bf A4}, 1693 (1990);
R. Dijkgraaf, E. Verlinde and H. Verlinde, ``Notes on
Topological String Theory and 2-D Quantum Gravity'', Lectures given at Spring
School on Strings and Quantum Gravity, Trieste, Italy, April 24-May 2,
1990 and at Cargese Workshop on Random Surfaces, Quantum Gravity and Strings,
Cargese, France, May 28-June 1, 1990.

\item{32.} W. Oevel and C. Rogers, Rev. Math. Phys. {\bf 5}, 299 (1993).

\item{33.} E. Date, M. Kashiwara, M. Jimbo and T. Miwa, in ``Nonlinear
Integrable Systems-Classical Theory and Quantum Theory'', ed. M. Jimbo and T.
Miwa (World Scientific, Singapore, 1983).

\item{34.} P. J. Olver , ``Applications of Lie Groups to Differential
Equations'', Graduate Texts in Mathematics, Vol. 107 (Springer, New York,
1986).

\item{35.} P. Mathieu, Lett. Math. Phys. {\bf 16}, 199 (1988).

\item{36.} W. Oevel and O. Ragnisco, Physica {\bf A161}, 181 (1989).

\item{37.} W. Oevel and Z. Popowicz, Comm. Math. Phys. {\bf 139}, 441 (1991).

\item{38.} J.M. Figueroa-O'Farril, J. Mas and E. Ramos, Leuven preprint
KUL-TF-91/19 (1991).

\item{39.} J. M. Figueroa-O'Farrill, J. Mas and E. Ramos, Rev. Math. Phys.
{\bf 3}, 479 (1991).

\item{40.} J. Barcelos-Neto and A. Das, J. Math. Phys. {\bf 33}, 2743 (1992).

\item{41.} G.H.M. Roelofs and P.H.M. Kersten, J. Math. Phys. {\bf 33}, 2185
(1992).

\item{42.} J. C. Brunelli and A. Das, J. Math. Phys. {\bf 36}, 268 (1995).

\item{43.} G. Wilson, Phys. Lett. {\bf A132}, 445 (1988).

\item{44.} F. Yu, Nucl. Phys. {\bf B375}, 173 (1992).

\item{45.}  J. Barcelos-Neto, S. Ghosh and S. Roy, J. Math. Phys. {\bf 36},
258 (1995).

\item{46.} Y. Ohta, J. Satsuma, D. Takahashi and T. Tokihiro, Progr. Theor.
Phys. Suppl. {\bf 94}, 210 (1988); K. Kiso, Progr. Theor. Phys. {\bf 83}, 1108
(1990).

\item{47.} J. Barcelos-Neto, A. Das, S. Panda and S. Roy, Phys. Lett.
{\bf B282}, 365 (1992).

\item{48.} P. H. M. Kersten, Phys. Lett. {\bf A134}, 25 (1988).

\item{49.} P. Dargis and P. Mathieu, Phys. Lett. {\bf A176}, 67 (1993).

\item{50.} E. Ramos, Mod. Phys. Lett. {\bf A9}, 3235 (1994).

\item{51.} D. Bernard and A. LeClair, Commun. Math. Phys. {\bf 142}, 99 (1989);
D. Bernard, ``An Introduction to Yangian Symmetries'', in Integrable
Quantum Field Theories, ed. L. Bonora et al., NATO ASI Series B: Physics vol.
310 (Plenum Press, New York, 1993).

\item{52.} J. Barcelos-Neto, A. Das, J. Maharana, Z. Phys. {\bf 30C}, 401
(1986);

\item{53.} N. J. Mackay, Phys. Lett. {\bf B281}, 90 (1992); erratum-ibid.
{\bf B308}, 444 (1993).

\item{54.} E. Abdalla, M. C. B. Abdalla, J. C. Brunelli and A. Zadra, Commun.
Math. Phys. {\bf 166}, 379 (1994).

\item{55.} T. Curtright and C. Zachos, Nucl. Phys. {\bf B402}, 604 (1993).

\end